\documentclass{aastex}          
\usepackage{spr-astr-addons}    
\usepackage{url}

\usepackage{color} 
\usepackage{hyperref}

\begin{document}

\title{On the Relation of the Lunar Recession \\and the Length-of-the-Day}
\shorttitle{Lunar occultations, tides, LLR, and LOD}
\shortauthors{Maeder and Gueorguiev}


\affil{\dag Corresponding author: \email{Andre.Maeder@UniGe.ch}}
\author{Andre Maeder\altaffilmark{1,\dag}}
\and 
\author{Vesselin G. Gueorguiev\altaffilmark{2,3}}
\affil{\email{Vesselin@MailAPS.org}}

\altaffiltext{1}{Geneva Observatory, University of Geneva, chemin des Maillettes 51, CH-1290 Sauverny, Switzerland}
\altaffiltext{2}{Institute for Advanced Physical Studies, 21 Montevideo Street, Sofia 1618, Bulgaria}
\altaffiltext{3}{Ronin Institute for Independent Scholarship, 127 Haddon Pl., Montclair, NJ 07043, USA}

\date{\today}

\begin{abstract}
We review the problem of the consistency between the observed values of the lunar recession from Lunar Laser Ranging (LLR)
and of the increase of the length-of-the-day (LOD). 
From observations of lunar occultations completed by recent  IERS  data, we derive  
a variation rate of the LOD equal to 1.09 ms/cy from
1680 to 2020, which compares well with McCarthy and Babcock (1986) and Sidorenkov (2005). This rate is 
lower than the mean  rate of 1.78 ms/cy derived by Stephenson et al. (2016) 
on the basis of   eclipses in the Antiquity and Middle Age.  The difference
in the two observed rates  starts at the epoch of a major  change in the data accuracy with telescopic observations.
The observed lunar recession appears  too large when compared to the tidal slowing down of the Earth 
  determined from eclipses in the Antiquity and Middle Age and even much more when determined
from lunar occultations and IERS data from 1680 to 2020.
With a proper account of the tidal effects and of the detailed studies 
on the atmospheric effects,  the  melting from icefields, the changes of the sea level, the 
glacial isostatic adjustment, and the core-mantle coupling, we conclude that the 
long-standing problem of the presence or absence  of a local cosmological expansion
is still an open question.

\end{abstract}

\maketitle

\tableofcontents{}

\section{Introduction}

Since 1970, the Lunar Laser Ranging (LLR) provides direct measurements of the 
Earth-Moon distance with an initial accuracy of the order of few dm 
and gradually over 5 decades arrived at the present few-mm level. 
The transition to the order of cm levels was attained from the 80s to the early 90s, see for example \citep{Williams16a}. 
The results indicate the presence of lunar recession of $38.30 \pm 0.09$ mm/yr. 
According to the formal definition from the International Earth Rotation and Reference Systems Service (IERS),
{\it{the excess revolution time}} (with respect to a day $\Delta T$ of 86400 SI seconds) is called the length-of-the-day (LOD).
We retain the standard notion to use LOD as the symbol for indicating the excess revolution time, 
while the change of the LOD is the derivative of $\Delta T$ with respect to time, it is usually determined in ms/cy.
The relation between the lunar recession and the change of the LOD depends on a number of effects, 
astronomical,  geophysical, climatic and atmospheric. Regarding the possibility of some additional
 contribution from cosmological expansion,  \citet{Dumin16} points out that  ``It is quite surprising that many
theorists believe that the possibility of local cosmological influences is strictly
prohibited just by the available observational data, while a lot of observers
believe that there are irrefutable theoretical proofs that Hubble-Lema\^{i}tre expansion is
absent at small scales.'' On both sides, the situation is still rich of uncertainties 
and this is what we try to study here.

The question whether  astrophysical systems, such as the solar system and  galaxies, 
follow the Hubble-Lema\^{i}tre expansion has stimulated a vast literature since the pioneer work of 
\citet{McVittie32,McVittie33} and the Einstein-Straus theorem \citep{Einstein45}.  
The presence of an expansion at smaller scales
has been considered as an open question by \citet{Bonnor00}.
The fact that the dark-energy dominates the matter-energy content of the Universe and that this energy
appears as driving the acceleration of expansion is reviving the interest in the question:
If dark energy is uniformly distributed in space would it not imply effects that may be present at small scales?
The Earth-Moon system occupies a particular place in this context, since there are direct accurate measurements
of the evolution of the distance in this two-body system.

In this respect, it is important to notice that the Hubble-Lema\^{\i}tre expansion rate of the Universe, in century (cy) units is
$H_0 \approx 70 \,({\rm km}/{\rm s})/{\rm Mpc}\approx0.7\cdot 10^{-8}\, {\rm cy}^{-1}$.
The  recession rate estimated using the average Earth--Moon distance $d = 384\,402$ km
that is increasing by  3.83 cm/yr results in $\dot{d}/d=0.999 \cdot 10^{-8}\, \textrm{cy}^{-1}$.
While these two numbers may be regarded as a puzzling coincidence, the mystery plot thickens with 
two independent astrometric and radiometric Cassini data \citep{Lainey} measurements 
of the mean recession speed of Titan from Saturn $v\approx 11.3{\rm ~cm/yr}$,
which when evaluated for the  average Titan-Saturn distance $D = 1\,221\,870$ km,
results in $\dot{D}/{D}\approx0.9\cdot 10^{-8}\, \textrm{cy}^{-1} $.
The possibility of a scale-dependent Hubble constant has also been raised already by 
\citet{Krizek15,Dumin18, Krizek21}. However, as it is well-known, 
the observed  recession is currently thought to be only produced by the tidal interaction in the Earth-Moon system,
which slows down the Earth rotation and thus produces an increase of the length-of-the-day. 
As this is the unique case where we have direct  high accuracy and continuous  measurements 
over centuries in the solar system, it is worth to examine  closely the present status of the problem 
in relation with the above cosmological question.

In Section \ref{lodeff}, we recall the basic relation between the LOD and the 
lunar recession imposed by the conservation of the total angular momentum in the classical  theory. 
Section \ref{cosmol} examines the various theoretical predictions in the cosmological context.
In Section \ref{phys}, the different  atmospheric and geophysical effects 
affecting the LOD are critically examined.  In Section \ref{LOD},
the data sources  on the LOD are collected and discussed, these are mainly the lunar and solar eclipses from the Antiquity
and Middle Age,  the lunar occultations of the past centuries and the data from the IERS. The analysis of the results is made 
in Section \ref{disc}. Section \ref{concl} gives the conclusions.

\section{The   LLR  and its tidal relation with the length-of-the-day (LOD)}       
\label{lodeff}

The Earth-Moon tidal coupling produces some lunar recession observable by the  Lunar Laser Ranging (LLR) experiment.   
The tidally distorted Earth rotates on its axis faster than the  Moon on its orbit.
Thus, the terrestrial deformation applies an additional forward attraction to the Moon, 
the angular momentum of which progressively increases at the expense of that of the Earth.
Thus, the  lunar recession is closely coupled to the slowing down of the Earth, {\it{i.e.}} the  increase of  the  LOD.
Many other effects also influence the LOD.
For example, the Earth's moment of inertia  changes due to the melting of polar icefields,  the variations of the sea level
and the  isostatic rebound following deglaciation.
There are exchanges of angular momentum  between the  solid Earth and atmosphere,  the core and the mantle.

\subsection{The Lunar Laser Ranging (LLR)}   \label{LLR}

The first retroreflector on the Moon was placed in 1969 by Apollo 11,
with 100 single reflectors with a diameter of 3.8 cm.
Successive improved reflectors were placed 
by the Apollo 14 and 15 flights  \citep{Dickey94}. 
The Apollo 15 retroreflector consists of 300 cube corners.   
With the reflector from the Lunokhod 2 rover (1973), 
a French array of 14 corner cubes of 11 cm on a Soviet rover, 
along with Lunokhod 1 which was lost but found in 2010 \citep{Lunokhod1}, 
this makes a total of 5 reflectors.
Short pulses of light are transmitted to the Moon and reflected back, the travel time being measured. 
The narrow initial beam is spread over an area of about 7 km diameter
on the Moon, while  the spot back on the Earth has a diameter of 20 km.
Thus,  only a  very small fraction  of the photons (about $10^{-21}$) of the initial beam finally 
reaches   the telescope  back on the Earth.
The first 15 years of the USA observations were conducted at McDonald Observatory, Texas.  
Observations were also performed in the Crimean Astrophysical Observatory  with
the retroreflectors installed on the Soviet rowers Lunokhod 1 and 2 launched in the early 70's, see \citet{Shiga10}.
The initial accuracy of distance measurements was 
25 cm, due to improvements from 1980 it went down to $2-3$ cm. In the 80's, 
new observing stations opened in Observatoire de la Côte d'Azur, France,  and
in Haleakala Observatory, Hawaii. The observations we mention below come from these observatories, as well as from 
Apache Point Observatory, New Mexico, and Matera, Italy.  More recently new observing stations were developed in 
Hartebeesthoek-South Africa, Yunnan-China (meter-level accuracy), and Wettzell-Germany (mm-level accuracy), 
which considerably improve the accuracy.
Between March 1970 and September 2015, there were 20138 ranges which led to an estimate of the 
lunar recession of  $38.30 \pm 0.09$  mm/yr
\citep{Williams16a,Williams16b}, a result that has not much changed since
about three decades, since  first determination by \citet{Christodoulidis88}.
The geophysical model by Williams et al. associated
to the determination of lunar orbital parameters
predicts a corresponding   increase of  the LOD of  2.395 ms/cy (millisecond per century).

\subsection{The LOD}

Since 1962 there are considerable improvements in the  determination of the LOD and  our  geodetic knowledge. 
This is  particularly due to the development of new techniques, in particular  the  space geodesy, SLR (Satellite Laser Ranging),
GNSS (Global Navigation Satellite System), DORIS (Doppler Orbitography and Radio positioning Integrated by Satellite)
and  VLBI (Very Long Base Interferometry). 
The International Earth Rotation and Reference Frame System (IERS)
(see the IERS URL to the Science/EarthRotation)
is the recognized world basis  for the study of   the fundamental parameters of the Earth motion
and orientation and the references frames.

Let us recall some key definitions.  The  time UT (Universal time)  is defined by 
two consecutive crossings of the meridian by the  Sun, this timescale depends on the Earth's rotation.
It is different  from  the time TT (Terrestrial time)  based 
on the SI second (International System) defined  by the  oscillations of the  cesium atom
in an atomic clock.  A day in TT scale is defined by  86400 SI seconds. The difference 
$\Delta T = TT - UT$ gives the accumulated difference over a given period of time. 
The  length-of-the-day (LOD), as already mentioned, is the 
the excess revolution time with respect to a day of 86400 SI seconds. 
The change of the LOD is the derivative of $\Delta T$ with respect to time, it is usually determined in ms/cy.
Fig. \ref{Iers} below illustrates the LOD from 1962 to 2020 as given by the IERS.
Given the already mentioned natural confusion due to the literal interpretation of the abbreviation LOD as 
Length of the Day, it should be stressed that it is a standard notion to use LOD as the symbol for indicating 
the excess revolution time, and as such, it is not anywhere close to 86400 SI seconds but more into ms range.

\begin{figure*}[h]
\centering 
\includegraphics[width=.75\textwidth]{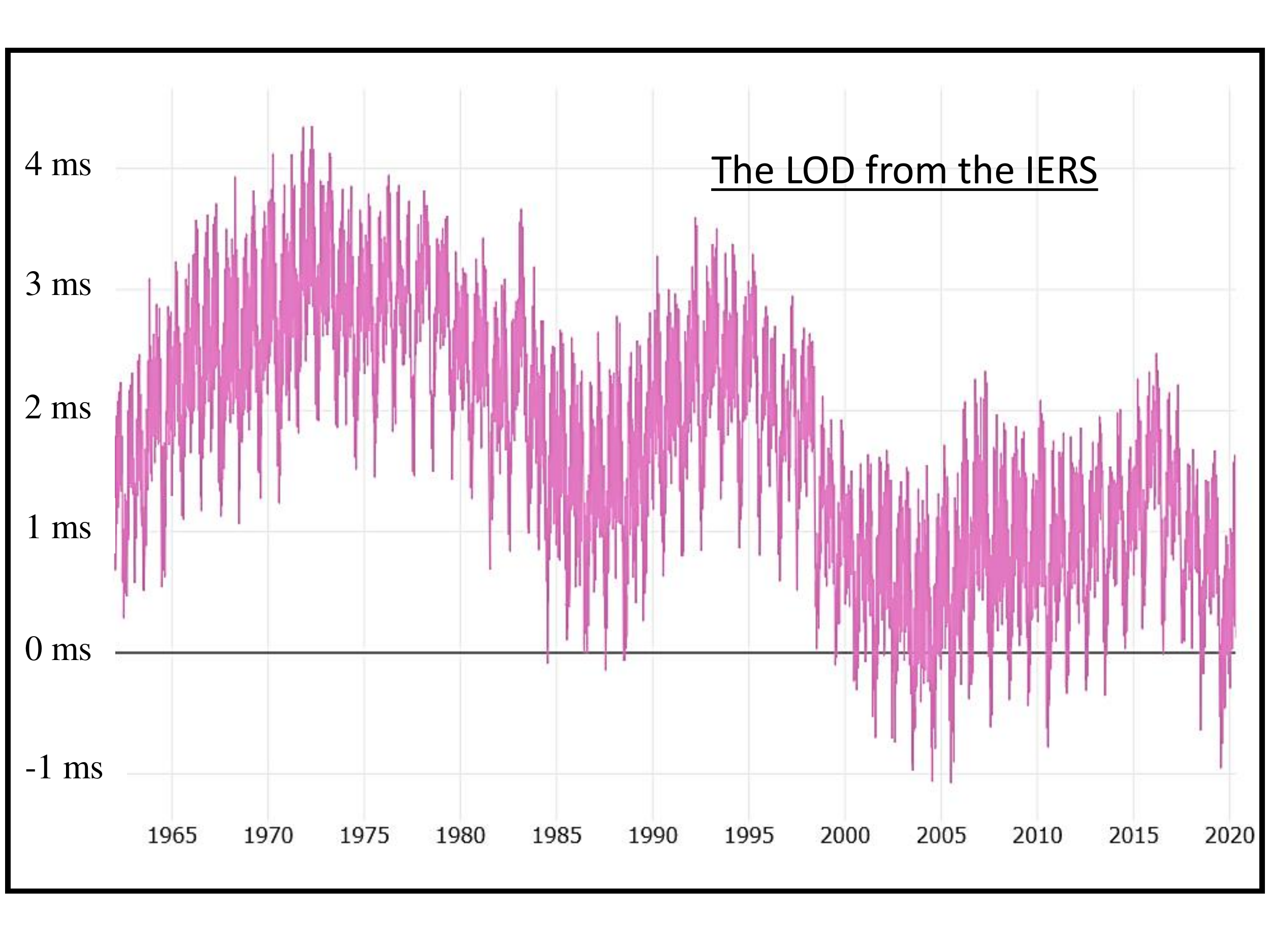}
\caption{The variations  of the  length of the day (on the left axis)  
in s as a function of time given in years from the International Earth Rotation Service  (IERS 2020).
The data cover the period of 1962 (with time obtained from atomic clocks) to the year 2020.
The IERS data shown above exhibits a negative slope of $-0.0001 \pm 8.29\times 10^{-7}$ ms/yr.
We see the  large seasonal variations and those over decades.}
\label{Iers} 
\end{figure*}

\subsection{Tidal effects}  \label{tide}

Tides are due  to the differential effects of gravitation on the various parts of a body. They depend on $R^{-3}$,
where   $R$ is the distance to  the external source   of gravitation generating the tide, see \citet{Tokieda13} for basics on tidal effects. 

The tides produced by the Moon and the Sun are in an intensity  ratio typically of  2.4 to 1. There are dozens of tidal waves of
different  periods  $P$  excited by the Moon and the Sun on the oceans, the Earth's crust and atmosphere \citep{Simon13}.
The  designation of the tidal waves was defined by Georges Darwin in 1883 and is still in use.
The main ones  are the semi-diurnal tides which have the largest intensity coefficients 
(Lunar mean M2 P= 12.42 h, solar mean S2 P=12.00 h, lunar main elliptic N2 P=12.66 h).  
The diurnal tides (Lunar principal O1 P=25.82 h, lunar declination K1 P= 23.93 h,
solar principal P1 P= 24.07 h) have a lower intensity, for example the ratio of O1 to M2 is 42\%. There are also the bimonthly 
(or fortnightly) tides excited by the Moon (Mf P=13.66 days) with an intensity of 17\% of M2. The lunar monthly tides
(Mm  P=27.55 days) have an intensity of 9\% of M2, the solar semi-annual (Ssa P=182.62 days)  an intensity of
8\% of M2, the solar annual (Sa P=365.24 days)   an intensity of 1.3\% of M2.
Moreover, it is to be noted that a tidal wave, as for example M2, can be resolved in a spectrum of multiple
nearby  frequencies  showing about a Gaussian  distribution of intensities  around the main component.
Concerning the effects on the  Earth-Moon distance, the main active tidal components are M2, O1 and N2, 
the semi-diurnal tides contribute to 88\% of the effect, the diurnal ones  to
13\%, while the Moon internal tides contributes to  1\% with the opposite sign \citep{Williams16a}.

The tidal waves   have complex figures on the sphere:
zonal, sectorial, and  tesseral  \citep{Simon13}.
The oceanic  and bodily tides are the most effective and also the main driving 
of the secular increase of the LOD and  lunar recession. 
The amplitudes of the LOD variations, due to the M2 semi-diurnal oceanic tides are of the order of  $\pm 1$ ms,
as  visible from Fig. \ref{Iers} from the IERS. 
Following  the semi-diurnal  tides, the fortnightly  oceanic and crustal
Mf tides which are zonal and symmetric  with a
period 13.66 day are  also quite effective. A theoretical model, which 
accounts for 14 years of altimeter data from satellites \citep{Ray12},
estimates that the amplitude of their effects on the LOD is about 0.36 ms,
a  significant source of noise in the variations of the LOD around the mean in the IERS data.
The oceanic and bodily tides  produce monthly Mm variations of the LOD
estimated to be about $\pm 0.30$ ms \citep{Sung-Ho16}.
Let us mention that the crustal tides may reach up to half a meter at the equator, their effects on the LOD have been
estimated  to reach 0.1 ms by \citet{Defraigne99}.

\subsection{Calculation of the Earth-Moon tidal coupling} \label{EarthM}

The full system of equations describing the changes in the rotation of an
elastic  axisymmetric body having a fluid core and equilibrium oceans
that is subject to small perturbing
excitation has been rediscussed by \citet{Gross09}. 
This system permits to study  the changes of the Earth rotation produced by  
small changes in relative angular momentum or in the Earth’s inertia tensor. 
Here, our aim is  to connect quantitatively 
in a simple approach the lunar recession and the tidal braking of the Earth  rotation.
The main effect in the Earth (E)--Moon (M) system can be estimated  by the angular momentum conservation
of the whole system  \citep{Dumin03}. This implies,
\begin{eqnarray}
\cos \varphi I_{\mathrm{E}} \Omega_{\mathrm{E}}
+ M_{\mathrm{M}}  R^2 \Omega_{\mathrm{M}} =  
\cos \varphi \sum_i I_i  \Omega_i 
+ M_{\mathrm{M}}  R^2 \Omega_{\mathrm{M}} \nonumber \\ 
= const.
\label{cons}
\end{eqnarray}
The index ``E'' refers to the global Earth, while the summation
is made on the various parts  {\it{i}} of the rotating Earth: mantle, atmosphere and  core.
The angle $\varphi$, considered the same for the various rotating  components, 
is the variable angle between the lunar orbital plane  and  the Earth equator.
$M_{\mathrm{M}} $ is the mass of the Moon and $\Omega_{\mathrm{M}}$ its
orbital angular velocity. $R$ is the mean distance between the Earth and Moon. 
We neglect the axial angular momentum of the Moon, since its mass 1.2 \% of that of the Earth and its axial rotation period (equal to its 
orbital period) is 27.3 days. Thus, the lunar axial angular momentum is a fraction of about $4 \cdot 10^{-4}$ of that of the Earth.
As the angle $\varphi$ varies over the relatively short period of the lunar nodal precession (18.60 yr), 
we consider below a mean value of this angle.
\begin{eqnarray}
\cos \varphi \, \left[\sum_i I_i \left( \frac{\partial}{\partial t} \Omega_i \right)_{\mathrm{I_i}}+ \sum_i
\Omega_i \,  \left(\frac{\partial I_i}{\partial t}\right)_{{\Omega_i}} \right]+\nonumber \\
+M_{\mathrm{M}}\frac{d}{dt} ( R^2 \Omega_{\mathrm{M}} )= 0 \,.
\label{dcons} 
\end{eqnarray}
Account  is given to the possible changes of  the moment of inertia.
The mean  orbital angular velocity of the Moon is
$ \Omega_{\mathrm{M}}=  G^{1/2} M_{\mathrm{E}}^{1/2} R^{-3/2}$, where $M_{\mathrm{E}}$ is the Earth's mass. The
previous equation becomes, if $T_{\mathrm{i}}= 2 \pi/ \Omega_{\mathrm{i}}$ is the period of the axial rotation of the i-component 
(a day for the mantle),
\begin{eqnarray}
2 \pi \cos \varphi \left[-\sum_i I_i \frac{1}{T_i^2} \left(\frac{\partial T_i}{\partial t}\right)_{\mathrm{I_i}}
+\sum_i \frac{1}{ T_i} \,\left(\frac{\partial I_i}{\partial t}\right)_{\mathrm{T_i}} \right]+ \\ \nonumber
G^{1/2} \,M_{\mathrm{M}}
\frac{M_{\mathrm{E}}^{1/2}}{2 \, R^{1/2}} \frac{dR}{dt} = 0 \, .
\label{basic}
\end{eqnarray}
This equation can be written as,
\begin{equation}
\frac{dR}{dt}\, = \, \sum_i \left[ k_i \, \left(\frac{\partial T_i}{\partial t}\right)_{\mathrm{I_i}}
- k'_i \left(\frac{\partial I_{\mathrm{i}}}{\partial t}\right)_{\mathrm{T_i}}\right]\, ,
\label{b2}
\end{equation}
with $k_i$ and $k'$ defined by
\begin{equation}
k_i  =  \frac{I_i}{T_i} k'_i \, \; \mathrm{and} \; \,
k'_i =  4 \pi \cos \varphi \sqrt{\frac{R}{T^2_{\mathrm{i}} G M^2_{\mathrm{M}}M_{\mathrm{E}}}}\,.
\label{ki}
\end{equation}
Eq. (\ref{b2}) expresses the lunar recession as a function of the change of rotation period of the Earth's components
and  the change of the moments of inertia.
If we  consider only the   tidal interaction between the Earth globally  and the Moon, 
we get similarly for a constant global moment of inertia $I_{\mathrm{E}}$,
\begin{equation}
\frac{dR}{dt} = \, k_{\mathrm{E}} \,  
\frac{dT_{\mathrm{E}}}{dt} \,, \;  \mathrm{with}\;
k_{\mathrm{E}} =4 \pi \cos \varphi \frac{R^{1/2} I_{\mathrm{E}}}
{T^2_{\mathrm{E}} G^{1/2}M_{\mathrm{M}}M^{1/2}_{\mathrm{E}}}.
\label{TR}
\end{equation}
Thus, $  \frac{dT_{\mathrm{E}}}{dt}$ is the Earth's slowing down due  to the tidal interaction.
We  mentioned above that a complex  model of  the tidal interaction   leads to an estimate of the increase of the LOD
of $  \frac{dT_{\mathrm{E}}}{dt} = 2.395$ ms/cy, corresponding to the observed  lunar recession  
of  $\frac{dR}{dt}=3.83$ cm/yr \citep{Williams16a,Williams16b}.
We adopt the following numerical values:
\begin{eqnarray}
M_{\mathrm{E}} = 5.973 \cdot 10^{27} \mathrm{g},  \quad  \quad R_{\mathrm{E}}= 6.371 \cdot 10^{8} \mathrm{cm}, \\ \nonumber
\; \; \quad M_{\mathrm{M}}= 7.342 \cdot 10^{25} \mathrm {g},\quad \quad R=  3.844 \cdot 10^{10} \mathrm{cm},\\ \nonumber
\quad I_{\mathrm{E}} = 0.331 \cdot  M_{\mathrm{E}}   R^2_{\mathrm{E}} =
8.0184 \cdot  10^{44} \mathrm{g \cdot cm}^2. \nonumber
\end{eqnarray}
The numerical coefficient 0.331  is obtained from precession data by  \citet{Williams94}.
The angle $\varphi$ varies between 18.16  and 28.72 degrees, thus we adopt a mean value of $\cos \varphi = 0.914$.
The coefficient $k_{\mathrm{E}}= 1.650 \cdot 10^5$ cm $\cdot$ s$^{-1 }$. 
For $k'_i$, taking $T_i \equiv T_{\mathrm{E}}$, we get  $ k'_i = 1.777 \cdot 10^{-35}$  g$^{-1}$ cm$^{-1}$.
With the appropriate units used in the study of the LLR, we write Eq.(\ref{TR}),
\begin{eqnarray}
\frac{dR}{d t}\left[\frac{\rm cm}{\rm yr}\right] \,=\, 1.650 \cdot 10^5 \left[\frac{\rm cm}{\rm s}\right] \;   
\frac{dT_{\mathrm{E}}}{dt}  \left[\frac{\rm s}{\rm yr}\right] \,.
\label{TRU}\, 
\end{eqnarray}
This expresses  the tidal  relation between  the  lunar recession  and the  increase
of  the axial  rotation period   $T_{\mathrm{E}}$ of the Earth, {\it{i.e.}} the LOD, in absence of other effects.
An estimate based on the above  expression (\ref{TRU}), 
applied to the observed Moon recession, gives  an increase of the LOD
of $2.32 \cdot 10^{-5}$ s $\cdot$ yr$^{-1}$, a value remarkably close   (at 3.1 \%) to that derived from a more 
detailed model.
This means that the angular momentum conservation effectively contains the resulting pull of the numerous tidal 
waves and that the approximations made have  a  limited incidence. 
Mechanical effects in the Earth-Moon system 
have an ``instantaneous'' action. Let us consider the Earth-Moon 
system as a two-body isolated system. 

The best way to assess the possible effect of the Hubble expansion (cf. Section \ref{cosmol}) 
is via the comparison of the observed lunar recession to the anticipated tidal effects. 
If there is  an additional cosmological contribution to the lunar recession, 
the total observed recession $\left(\frac{dR}{dt}\right)_{\mathrm{obs}}$
is the sum of two contributions, the effectively tidal one $\left(\frac{dR}{dt}\right)_{\mathrm{tdeff}}$
and the cosmological one  $\left(\frac{dR}{dt}\right)_{\mathrm{cosm}}$. Thus, one has
\begin{equation}
\left(\frac{dR}{dt}\right)_{\mathrm{tdeff}} = \left(\frac{dR}{dt}\right)_{\mathrm{obs}}
-\left(\frac{dR}{dt}\right)_{\mathrm{cosm}}\,.
\label{cosm}
\end{equation}
This expression will be used in establishing Table \ref{Table1} below.

Equation (\ref{TR}) which expresses  the global tidal coupling,  
can be written as $\left( \frac{dT_{\mathrm{E}}}{dt}\right)_{\mathrm{tdeff}} =
\frac{1}{k_{\mathrm{E}}} \,\left(\frac{dR}{dt}\right)_{\mathrm{tdeff}}$
that is corresponding  to the effective  tidal change  of the Earth's rotation period  and
thus is given by  Eq.(\ref{b2}) and leads to:
\begin{equation}
k_{\mathrm{E}} \left(\frac{dT_{\mathrm{E}}}{dt}\right)_{\mathrm{tdeff}} \, =
\, \sum_i \left[ k_i \, \left(\frac{\partial T_i}{\partial t}\right)_{\mathrm{I_i}}
- k'_i \left(\frac{\partial I_{\mathrm{i}}}{\partial t}\right)_{\mathrm{T_i}}\right]\, .
\label{b3}
\end{equation}
The above equation  allows us to  distinguish between the effects producing  exchanges 
of angular momentum from those which modify the Earth's moment of inertia. The atmospheric motions
(circulation currents, zonal winds, etc.) produce exchanges of angular momentum between the
atmosphere and the mantle,  other exchanges also occur between the mantle and the core. Thus, we consider below
in a simplified way 
three ``i'' components of the Earth: the atmosphere (atm), the mantle (mtl) and the core (cr), 
which may all experience  changes of rotation periods.
Thus, the above expression can be rewritten
\begin{eqnarray}
\frac{dT_{\mathrm{mtl}}}{dt} =  \frac{k_{\mathrm{E}}}{k_{\mathrm{mtl}}}
\left(\frac{dT_{\mathrm{E}}}{dt}\right)_{\mathrm{tdeff}} - \frac{k_{\mathrm{atm}}}{k_{\mathrm{mtl}}}
\frac{dT_{\mathrm{atm}}}{dt} \nonumber \\
- \frac{k_{\mathrm{cr}}}{k_{\mathrm{mtl}}}
\frac{dT_{\mathrm{cr}}}{dt} +  \sum _i\frac{k'_i}{k_{\mathrm{mtl}}}
\frac{dI_{\mathrm{i}}}{dt}\, ,
\label{b4}
\end{eqnarray}
with the condition that for each derivative the other factors are considered as constant.
The   ratios of the k-terms express the  weighting  according to the moments of inertia,
and assuming an equality of the periods in the various k-factors, one has:
\begin{eqnarray}
\frac{k_{\mathrm{E}}}{k_{\mathrm{mtl}}} \approx  \frac{I_{\mathrm{E}}}{I_{\mathrm{mtl}}} \approx 1.013, \,
\frac{k_{\mathrm{atm}}}{k_{\mathrm{mtl}}} \approx  \frac{I_{\mathrm{atm}}}{I_{\mathrm{mtl}}} \approx  1.7\cdot 10^{-6}, \nonumber\\
\frac{k_{\mathrm{cr}}}{k_{\mathrm{mtl}}} \approx  \frac{I_{\mathrm{cr}}}{I_{\mathrm{mtl}}} \approx  0.013.
\label{ik}
\end{eqnarray}
For  estimates, we  use   a core mass equal to 32.5 \% of that of the Earth with a radius of 1217.5 km \citep{Kennett95},
with a numerical coefficient of 0.36 to obtain $I_{\mathrm{cr}}$. 
We have $I_{\mathrm{mtl}}=I_{\mathrm{E}}-I_{\mathrm{cr}}$
and a  mass of the atmosphere of $ 5.15\cdot 10^{21}$ g.
The term on the left in Eq. (\ref{b4}), giving the change of the rotation period of the mantle,  corresponds to the observed 
change $\dot{\tau}_{\mathrm{obs}}$ of the LOD, which is   expressed
in ms/cy. The first term on the right is the change $\dot{\tau}_{\mathrm{ tdeff}}$ 
of the LOD corresponding to the effective lunar recession, the cosmological effect being subtracted, if anyone. 
The correspondence is based on the fact that  a lunar recession of 
3.83 cm/yr is related to a change of  the LOD of 2.395  ms/cy. 
The second and third term on the right are respectively 
the atmospheric  $\dot{\tau}_{\mathrm{atm}}$  and core $\dot{\tau}_{\mathrm{cr}}$  contributions.
They allow us to express the effects of the exchange between the solid Earth with the atmosphere and the core.
Thus, if  $T_{\mathrm{atm}}$ or $T_{\mathrm{cr}}$ increases, the angular momentum conservation
implies that the observed  change of the LOD should be reduced.
Any increase of the moment of inertia of the Earth would  increase the LOD. Changes occur in particular  by  melting
of high latitude icefields and by the resulting changes of the sea level,  as well by
the isostatic rebound also called Global Isostatic Adjustment (GIA). 
Finally, we may write 
\begin{equation}
\dot{\tau}_{\mathrm{obs}} \, = \,  \dot{\tau}_{\mathrm{tdeff}} +  \dot{\tau}_{\mathrm{atm}}+  
\dot{\tau}_{\mathrm{cr}} +  \dot{\tau}_{\mathrm{ice}}
+  \dot{\tau}_{\mathrm{sea}} +  \dot{\tau}_{\mathrm{GIA}} \, ,
\label{sum}
\end{equation}
where each term represents a specific contribution to the time variation of the LOD expressed in ms/cy:
terms $\dot{\tau}_{\mathrm{atm}}$ and $  \dot{\tau}_{\mathrm{cr}}$ for the exchanges of the mantle with atmosphere and core,  
the next ones  for the change of the moment of inertia due to  ice melting,  change of the sea level and 
glacial isostatic adjustment respectively.
Thus, we verify that the angular momentum conservation allows the addition, properly weighted by the moment of inertia, of the various 
characteristic times.  An expression of that sort has been written by \citet{Munk02}. Now, the great challenge
is to get the numerical values of  the  terms contributing to this equation.

The tidal effects can also be theoretically calculated from the predicted tidal distortions of the Earth's atmosphere,
oceans, and mantle. There appear  some problem here. Globally, these distortions would lead to a much
smaller tidal effects than the one anticipated from the lunar recession. They would only account for
about 55\% of the observed lunar recession of 3.83 cm/yr, as emphasized by \citet{Krizek15} on the basis of the 
several studies. This was  first mentioned by \citet{Flandern75}. 
Also, the disagreement between the rate of lunar recession and the LOD variation was attributed to a possible
cosmological expansion by \citet{Dumin01}. 
The tidally deformed figure of the Earth has also been discussed by 
\citet{Novotny98} and further  analysed in the geophysical studies of the history of the  Earth's rotation by \citet{Peltier07}.

\subsection{The Moon's orbital motion}  \label{Moon}

\begin{figure*}[h]
\centering 
\includegraphics[width=.75\textwidth]{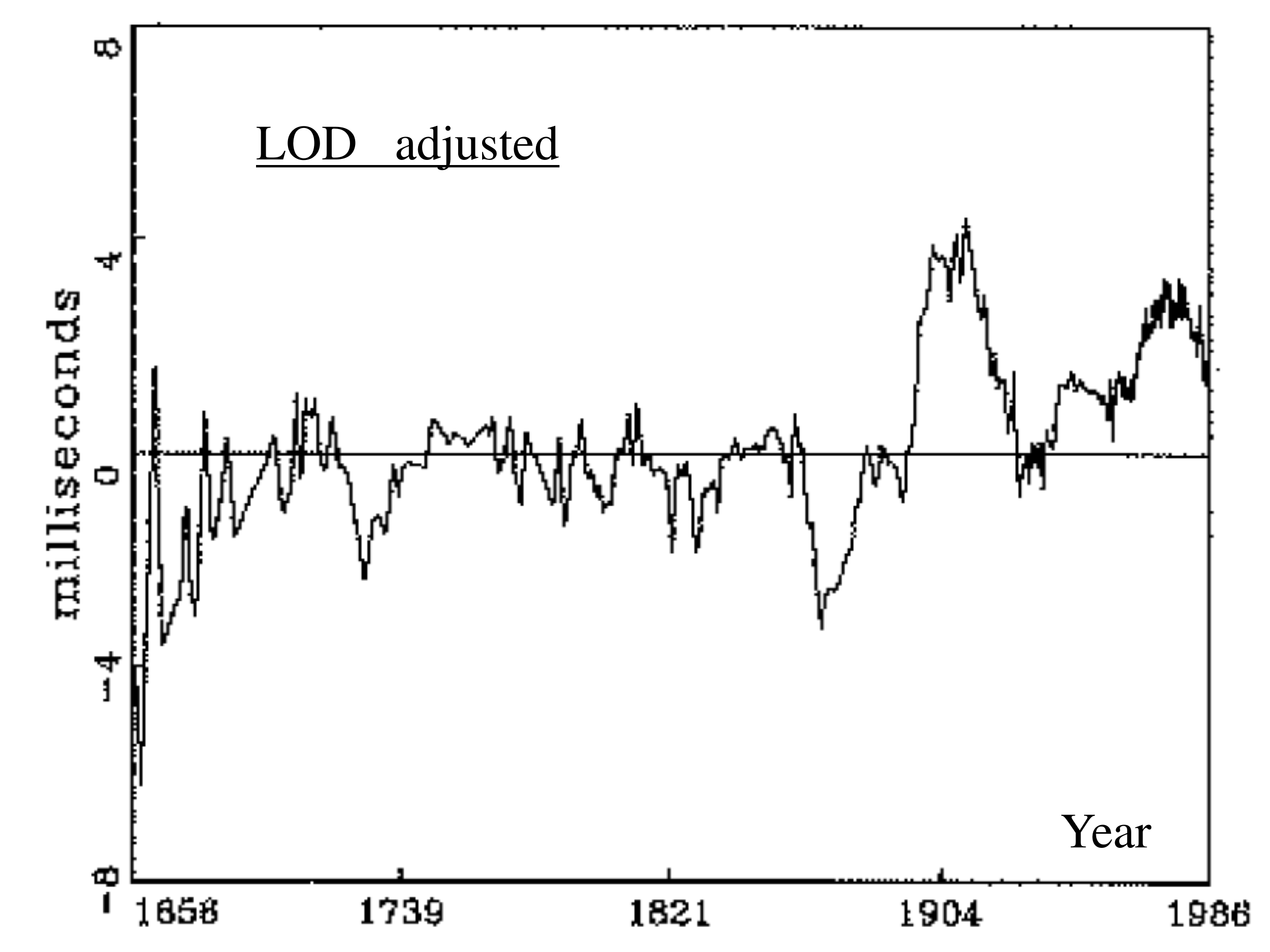}
\caption{Graphic representation of the data by \citet{McCarthy86} on the variations of the LOD since 1656.
The values have been adjusted by these authors for the mean decrease of the angular velocity of the Earth 
given in Eq. (\ref{sec}). This way the mean trend of +0.73 ms/cy 
appears to be at zero - the horizontal line of the Figure from \citet{McCarthy86}.}
\label{McC} 
\end{figure*}

The assumption of a circular orbital motion of the Moon is a crude assumption.
The orbit is an ellipse with eccentricity of 0.055, so that the distance between the 
perigee and apogee varies by about 43 000 km with a synodic period of 29.53 days.
A number of other effects in the lunar motion  intervene: the apsidal precession 
(change of the orientation of the semi-major axis of the elliptic
lunar orbit with a period of 8.85 yr);
the nodal  precession (related to the precession of the orbital plane of the
Moon), as a consequence the angle $\varphi$     
between the lunar orbital plane  and  the Earth equator varies with a period 18.60 yr, as seen above.
Some of the secondary oscillations of the LOD in Figs. \ref{Iers}, \ref{McC}, and  \ref{AV1700}
may be due to the apsidal and nodal precession. There are also  the axial precession of the 
Moon, its physical libration and  nutation  which have more limited effects \citep{Dehant09}.
The precession of the Earth is acting on a much longer timescale of 26 000 yr.

\section{Cosmology and  expansion at small scales}  \label{cosmol}

The question arises whether there are also some small additional effects of the global cosmological expansion  locally at small scales.
Following \citet{McVittie33}, most authors  were studying  this problem  with a metric  combining
an exterior Schwarzschild solution of a mass concentration,
and a FLRW metric representing the expanding Universe. For a review on the attempts
to estimate the influence of the cosmological expansion on local systems, see  \citet{Carrera10}. Most authors
generally concluded to a  negligible effect, see \citet{Cooperstock95}  and \citet{Sereno07}. \citet{Anderson95} found
some expansion of small systems, but at a negligible rate.
\citet{Price05}  proposed ``the all or nothing'' expansion,  a work  discussed by \citet{Faraoni07}, who pointed out 
that it  applies to a de Sitter background, towards which the  Universe is tending.
The absence of an effect is also more a postulate, not to say  a dogma, in some major references, see \citet{Misner73}. 
These authors also postulated that Hubble expansion must be truncated at some sufficiently small scale, because otherwise
a unit of length could not be defined by which the expansion can be measured. 
However, there are a few different voices.  \citet{Bonnor00} considered that:
``The scale at which the cosmic expansion begins to affect objects seems
to be an open question. Intuitively it is hard to think of objects held together
by strong forces, such as those of the electric field, as taking part
in the Hubble expansion. However, cosmologists often say the expansion
is that of space itself. From this point of view it would not be so counter intuitive
if ordinary bodies expanded with the universe.'' 
Ordinary bodies, like crystals, atoms, and nuclei do not expand due to the predominance of
non-gravitational forces like electromagnetic and strong force.
We should deal only with free bodies that interact gravitationally.

In the  theorem  by \citet{Einstein45},  the authors consider  a central mass 
$M= \frac{4 \pi \varrho R^3}{3}$ made by the concentration
of matter with a mean density $\varrho$ from   a spherical cavity of radius $R$, embedded in an expanding Universe. 
The theorem states that there is no expansion in the empty cavity, while
beyond it,  including  the limit $R$,  the expansion is present, operating according to the Hubble law.
In a simplified way, the  argument means that the limit between expansion and no expansion occurs
when the kinetic energy  $ (1/2) \upsilon^2= (1/2) (H \,R)^2$ of the expansion
is equal  to the local potential energy  $GM/r$ for a spherical geometry (taking the Virial expression 
or not makes no big difference).
This corresponds to $H= \sqrt{\frac{8\,\pi}{3} G {\varrho}}$,
the standard expression of the Hubble rate in Friedman cosmology  for a flat Universe. 
Thus, inside the cavity, 
no expansion is predicted.
This is the global sense of the Einstein-Straus theorem and of  various approaches. 
The fact that there is an abrupt change  from no expansion to full expansion when crossing radius $R$, while gravity 
is a continuously varying function, is very  disconcerting, at least for the present authors. 

Nowadays the case of the Einstein-Straus theorem is even more problematic, since
the Universe is dominated by  dark energy, which represents about 70\% of the matter-energy
and is the driving  force of the accelerated expansion. 
As dark energy   is the same  everywhere,  
the conditions of  the Einstein-Straus theorem are never met:  there should be 
no empty cavity free from mass-energy, and thus without expansion in the $\Lambda$CDM model, 
for more in depth discussion and critics of no-expansion claims see the original work by \cite{Dumin05}.

The  effects of the cosmological constant $\Lambda$, which expresses
the energy-density of the vacuum, have been considered by 
\citet{Balbinot88}, \citet{Klioner05} and  \citet{Sereno07}.
The conclusions were that there is some expansion, but  the effects were found
insignificant, for example \citet{Sereno07} estimate an expansion of 
the Earth-Sun distance with  a rate of $10^{-21}$ m/yr. 
There is a noticeable  exception with \citet{Dumin16,Dumin20}, who applies a new metric 
inspired from the original Kottler metric \citep{Kottler18}, which accounts for 
a central mass (cf. Schwarzschild metric) and  the $\Lambda$-term (cf. de Sitter metric). 
Dumin makes some further transformation of
the metric to  account for expansion. He obtains outwards spiraling orbits and conclude to the existence of 
a slightly smaller local Hubble expansion than the global one.
Thus, on the whole the conclusions of standard models are generally in favor of  a negligible effect 
of the Hubble expansion in planetary systems.
Nevertheless, important divergent claims are also expressed \citep{Bonnor00,Dumin16}.

The work of \citet{Klioner05} concludes that cosmological effects for solar system bodies 
are negligible over the age of the Universe. 
However, their imbedding of the Barycentric Celestial Reference System 
into a cosmological background considers only 
a central perturbation potential. While such a perturbation due to a 
central potential could be used to probe dark matter effects in the solar system,
it seems inadequate for probing a cosmic expansion effects that should not have a
central point. The inadequacy of central potentials for understanding very low
Newtonian gravity limits has been discussed previously by \citet{inMNRAS20}.

Several  non-standard cosmological models are exploring new research lines, stimulated  by the
fact that  about 95\% of the mass-energy in the Universe lies in  unknown dark components.
In the framework of scale invariant models, a general scale invariant field equation 
has been obtained \citep{Canu77} in the line of developments made by \citet{Dirac73}.
The scale invariant vacuum (SIV) theory  rests on the specific hypothesis 
that the macroscopic empty space is invariant  to a scale transformation $ds'=\lambda(t) \, ds$,
a property present in Maxwell's equation of electrodynamics,
as well as in General Relativity in absence of a cosmological constant, see review by \citet{MaedGueor20}.  
Cosmological equations have been obtained leading to accelerated Universe models \citep{Maeder17a}.
The weak field approximation of the geodesic equation \citep{Dirac73,BouvierM78} 
allows the study of the two-body problem,  which predicts that the semi-major axis $a$ increases like $\dot{a}/a = 1/t$, 
where $t$ is the time in the timescale of some cosmological models, thus the rate of expansion of a two-body system
is not necessarily the observed Hubble expansion rate, although it is likely of the same order of magnitude.
While the MOND theory \citep{Milgrom83,Milgrom19}
considers  a dilatation invariance with a constant scale factor,
the SIV theory assumes a scale factor  dependent  on time. 
The original weak-field approximation in the SIV utilized the assumption of isotropic and homogeneous space
for the derivation of the terms beyond the usual Newtonian equations \citep{Dirac73,BouvierM78}. 
The result is a velocity dependent extra term  $\kappa_0 \vec{v}$ with $\kappa_0=1/t$. 
Recently, similar term was derived as a result of un-proper time parametrization \citep{sym13030379}.
\citet{Krizek15} are emphasizing  that there is no reason that the dark energy responsible
for the accelerated expansion does not manifest itself in galaxies and in the Solar System as well. 
They examine a number 
of astrophysical problems, including the lunar recession, 
from which they suggest that even bound systems expand at a rate of the order of 
the Hubble expansion.

There  are several other  Extended Theories of Gravity, 
where some modifications of the dynamics are predicted.
The so-called $f(R)$ theories consider a gravity Lagrangian,  modified 
by including higher order terms depending on curvature $R$ 
\citep{Capozziello06,Capozziello07}.
The deformations act like a force that deviates the test particles.
The geodesics and general field equations  are consistent with those by \citet{Canu77}.
\citet{Sola15} consider a dynamical vacuum, 
with $\Lambda(H)= C_0+ C_{\mathrm{H }} H^2 + C_{\mathrm{\dot{H}}} \dot{H}$,
where the time dependence of 
$\Lambda$ improves a number of cosmological tests \citep{Sola15,Sola17}.
\citet{Meierovich12} uses a longitudinal vector field  
to represent the effects of the dark components  in the framework of GR.
In this respect, there are  some similarities with the SIV. The analysis below 
is independent of any cosmological models.

\section{Changes of the LOD due to various  terrestrial effects}   \label{phys}

The richness of the terrestrial effects influencing the LOD as they appear in Eq. (\ref{sum}) is impressive. 
Each one corresponds to a major current  research field. Here we attempt to briefly report on the main effects, 
their timescales and amplitudes.

\subsection{Atmospheric effects on the Earth's rotation and the LOD}  \label{atm}

The atmosphere  has an influence on  the Earth's spin rotation, by the exchanges of angular momentum between the solid
Earth and the atmosphere. The oscillation amplitudes grow exponentially with altitude
in the rarefied gas.    The exchanges are generated by atmospheric tides
and also by winds and global circulation currents in the atmosphere. The atmospheric effects on the LOD have 
a limited amplitude. Globally, the atmosphere super-rotates with an average velocity of 7 m/s with respect to the solid Earth.
If it would stop super-rotating and just co-rotate with the Earth, the  LOD would change by 3 ms 
accounting for the respective moments of inertia \citep{Gross09}.
This means that the large cyclic variations observed in the LOD (the so-called decadal  and the 1500 yr-oscillations)
cannot be due to exchanges between  the Earth's mantle and atmosphere. 
The   timescales for these exchanges are ranging
from half a day to a few  years, for a recent review see \citet{Sung-Ho16}.

Let us start by the short timescales.  The intense  semi-diurnal atmospheric tide  S2
(the major solar tidal component) has a  peculiar effect.
While the  oceanic lunar and solar  tides are excited by the gravitational differential pull,
the semi-diurnal atmospheric tides are first  excited by the Sun heating.
The maximum atmospheric pressure occurs in general  around 10 a.m.
local time,  {\it{i.e.}} before the Sun reaches the zenith, thus the solar attraction
exerts an additional pull on the present excess of atmospheric matter.
This means that  the  main component of the solar atmospheric tides has   an effect opposed to that of the Moon 
by accelerating the Earth's spin rotation, an effect first found by   \citet{Munk60}.
The energy exchange due to this effect has been re-estimated
to be about 4.5\%  of that of the  tidal energy dissipation of oceans and mantle \citep{Sung-Ho16}.
Since the energy of the rotation depends on the square of the period $T_{\mathrm{E}}$, 
the effect on the LOD is about 2.3\% of that due to the total tidal energy dissipation. 
This means that the semi-diurnal solar tide accelerates the Earth  rotation by something like 0.05 ms/cy.  
This small effect is often not accounted for.

The atmospheric effects have been  studied  by a number of authors over the last decades, 
see for example \citet{Eubanks85}, \citet{Hay16} and  \citet{Sung-Ho16}. Let us follow these last authors, 
who express the equations for the effects of  the pressure distribution on tides and make a quantitative analysis of the various terms.
They  have  also used  complete datasets of 6 hr interval global atmospheric pressure and wind velocity since 1980 
from the ECMWF (The European Centre for Medium-Range Weather Forecasts). On this basis,  they have performed 
comparisons between the meteorological  changes  and the variations  of the LOD, as  obtained from  the IERS website.
There is a strong connection between the change of the atmospheric angular momentum and the LOD, which results 
from surface friction by the winds, particularly the zonal winds, with a characteristic timescale of about 7 days \citep{Hay16}.
\citet{Sung-Ho16} confirm the seasonal variations of the LOD, which are due to changes in the global atmospheric circulation.  
 These effects, accompanied by polar drifts, are relatively large, 
the LOD is shorter from late June to early September by -0.65 ms, 
and longer by +0.65 ms from Winter to Spring time \citep{Sung-Ho16}, see also Fig. \ref{Iers}.

The atmospheric interactions also have timescales of years, as in the case of El-Ni\~{n}o--Southern--Oscillation (ENSO).  
\citet{Gross16} and \citet{Schindelegger17}  estimate that  about 30\% of the time-averaged rotation rate 
of the atmospheric contributions may be linked in some way to ENSO.
According to \citet{Sidorenkov97}, most of the periods in the ENSO
component spectrum (6, 3.6, (2.8), 2.4 yr), except for 2.8, 
are  multiple of the Chandler period of 1.2 yr, 
(the Chandler wobble is a small nutation of the Earth's rotation axis of 9 m amplitude).
Most fluctuations of the LOD in range of 40 to 700 days are due to atmospheric interactions \citet{Eubanks85}.
On the longer term, there is an equilibrium    between the atmospheric pick-up of angular momentum
at low and high latitudes and its transfer  to the Earth by West winds at middle latitudes \citep{Hay16}. \\

\noindent
{\it{Summary on atmospheric effects on the LOD:}}
On a timescale of more than three centuries, as we are considering in this study, the  positive
and negative atmospheric effects on the change of the LOD  largely cancel each other   on much shorter timescales. 
Thus, we must only account  for  the acceleration of the Earth rotation due to the semi-diurnal solar atmospheric tides  S2,
for  which   a contribution  of -0.05 ms/cy to the LOD was estimated here.

\subsection{Effects of ice melting and  change of the sea level}  \label{sealev}

The melting of glaciers and polar ice-fields and  the resulting increase of the sea level   have sizable consequences for  
the  Earth’s spin rotation rate, as well as for the polar wander of the rotation axis,  see review by  \citet{Mitrovica11}. 
The case  of polar  ice-fields is particularly critical at the present time, their strong melting since the year 1990
leads to an increase of the sea level over the world. Before 1990, the increase of sea level is mainly due to the
melting of glaciers outside from the polar caps.
Melting implies   a global motion of water away from the polar axis, 
producing  a growth of the Earth's moment of inertia, and thus an increase of the LOD. 
In a  pioneer work, \citet{Munk60}  estimated the relation between the changes of polar ice caps,  the sea level
and the length-of-the-day:  globally, a  1 cm eustatic rise  (change of the global ocean mass)    
of the sea level is leading to  a change of  the LOD equal to +0.1 ms, see also \citet{Munk02}

In the context of global warming,  an important melting produces a rapid increase of the 
sea level.  This is true nowadays, but the reality for the 20$^{th}$ century is rather  different.
The global balance between the ice accumulation in polar regions
and calving was about zero as suggested in many works by
\citet{Radok87}, \citet{Trupin93}, \citet{Johnston99}, \citet{Munk02} and the report 
of the IPCC 2001.  The  polar studies by \citet{Radok87} 
showed that the ice sheets of Antarctica and Greenland (the two major icefields) are in equilibrium
between accumulation and discharge, so that the net change
is  about zero.
Gridded estimates of ice accumulation on the Antarctic  showed an  
accumulation of ice of 4.5 cm/yr in the center of the continent
to about 50 cm/yr at the edge. On  Greenland there was evidences of a thickening in the years 1950-1970.
Thus, these two major polar icefields could even have   contributed to a reduction 
of the Earth's moment of inertia in the 20$^{th}$ century \citep{Trupin93}.
\citet{Johnston99}, in a study on the change of the oblateness term in the expression of the Earth potential,
also considered  the ice  accumulation and ablation from Antarctica and Greenland  to be about 
equivalent within $\pm 10 \%$. This result for these two sites  is also confirmed  by  the 2001 report of
the IPCC, which gives a value of -2 cm/cy for water storage 
and an effect of +3 cm/cy for ice melting, in equivalent effect on the sea level
(the estimates for Greenland are 0.5 ($\pm 0.5$) mm/yr and for Antarctica 0.5 ($\pm 1.0$) mm/yr).
Even more constraining,  a rise of sea level of 0.7 ($\pm 0.1$) mm/yr, due melting in polar regions 
for the period 1900-1990,  would lead to  disagreement with Lageos I
satellite determinations of the geopotential according to \citet{Mitrovica15}.
The change of the sea level  for the period 1900-1990  appears  essentially not due to
mass loss from polar icefields (up to a maximum of 0.2 mm/yr), but to the melting of other glaciers and water thermal dilatation
\citep{Mitrovica15}.
Year 1990  marks the start of
a high mass flux from polar regions,  as well as an increase of the fast melting of  mountain glaciers.

The eustatic contribution was estimated in 2001  by the Intergovernmental Panel on Climate  Changes 
(IPCC)   to correspond to 6 cm/cy, to which  a thermal dilatation effect of 3 cm/cy must be added \citep{Munk02}. 
Often previous estimates  suggested
a rise of the sea level of 1.5 to 2.0 mm/yr during most the 20$^{th}$ century. However, such high
values  lead  to serious misfit with geodesic observations.
More recently,  as reported by \citet{Mitrovica15}, the Fifth Assessment Report of the IPCC  (2014) estimates 
the sea level changes over the period 1900-1990 to come from three contributions:  
0.7 ($\pm 0.1$) mm/yr from glaciers (including those at the periphery of the Greenland Ice Sheet), 
0.4 ($\pm 0.1$) mm/yr from thermal expansion, and
- 0.11 ($\pm 0.05$) mm/yr from anthropogenic 
storage of water on land (we note that these values are concordant, but slightly higher than
those of the 2001 IPCC report for the same period). 
This was in good agreement with the results by \citet{Hay15} who estimate that over the period 1900-1990, 
before the beginning of a  rapid  general melting, the increase of the sea level, 
including the thermal dilatation (about 0.4 mm/yr), amounted to 1.2 ($\pm 0.2$) mm/yr, 
without significant contributions from the polar icefields over the period considered. 
If we adopt the correspondence established by \citet{Munk02} for {\it{polar regions}}, the eustatic change 
would imply  an increase of the LOD of about 0.7 to 0.8 ms per century. 
However, the rise of 0.7 to 0.8  mm/yr   originates from mountain glaciers
(in general at relatively high latitudes and on the coasts of Greenland),  
so that the variation of the LOD due to ice melting and change of the sea level
should be somehow smaller then 0.7:
\begin{equation}
\dot{\tau}_{\mathrm{ice+sea}}  \,< \,  0.7\, \mathrm{ms/cy}\,.
\label{s07}
\end{equation}

Fortunately, there is another possible source of information.
Satellite Laser Ranging (SLR) from low orbit satellite data provides information on the variation of the Earth's 
moment of inertia and oblateness. 
The Lageos I satellite range data firstly  revealed  significant changes  of  the degree -2 zonal harmonic  $J_2$ in the development  of the gravitational geopotential in spherical harmonics  \citep{Yoder83}, sensitive to the oblateness and moment of inertia. 
One has the following relation between the change of $J_2$ and $(\dot{\Omega}/\Omega)$ \citep{Mitrovica15},
\begin{equation}
\dot{J}_2 \;  \left[10^{-11} \cdot \mathrm{yr}^{-1}\right] \simeq  2 \, \frac{\dot{\Omega}}
{\Omega} \;  \left[10^{-11} \cdot \mathrm{yr}^{-1}\right] \, .
\label{equ1}
\end{equation}
A change of $-1$ ms of the LOD corresponds to the following value for  $(\dot{\Omega}/\Omega)$,
\begin{equation}
\Delta \mathrm{LOD} \; \mathrm{of} \; -1 \, \mathrm{ms}  \; \Leftrightarrow  \; 11.6 \cdot 10^{-11}
\mathrm{yr}^{-1}  \; \mathrm{for} \;  (\dot{\Omega}/\Omega)\,.
\label{equ2}
\end{equation}
The result for 5.5 yr of SLR was
$\dot{J}_2= 3\cdot 10^{-11}$ yr$^{-1}$, it 
corresponds  according to the above relations  to a change of the LOD of -0.51 ms/cy. 
For 10 year SLR from Lageos I  between
1980 and 1989, a value of  $\dot{J}_2= 2.6 \cdot 10^{-11}$ yr$^{-1}$
was found by \citet{Nerem93}, corresponding to a change of the LOD
of -0.45 ms/cy. 
Other  results were obtained from SLR  from 8 satellites from 1976
to 1995 by \citet{Cheng97}, who found  
$\dot{J}_2= 2.7 (\pm 0.4)\cdot 10^{-11}$ yr$^{-1}$
implying   a LOD variation of -0.47 ms/cy. More recent data of  satellite laser
ranging by \citet{Cheng13}  even allowed  them to measure the second derivative of the $J_2$ parameter. 
However, calculation of the second derivative from inexact data is an ill-conditioned problem which is numerically very unstable.
For the period 1976-1995, they find a value of $\dot{J}_2= 2.8 (\pm 0.3)\cdot 10^{-11}$ yr$^{-1}$
implying   a LOD variation   $\dot{\tau}_{\mathrm{J2}}= -0.48 \,(\pm 0.05)$  ms/cy. The convergence of all these values is
excellent. In view of the acceleration of
the motion, it seems preferable to keep the slightly lower value of -0.45 ms/cy  from 1980 to 1989.

The variation of $\dot{J}_2$ implies a change of the LOD, here called  $\dot{\tau}_{\mathrm{J2}}$,  
which accumulates the effects on the moment of inertia due to changes of
the sea level resulting from ice melting and those due to the global isostatic adjustment (GIA),
\begin{equation}
\dot{\tau}_{\mathrm{J2}}\, =\, \dot{\tau}_{\mathrm{GIA}}+ \dot{\tau}_{\mathrm{ice+sea}} \, .
\label{sumj2}
\end{equation}
In Section \ref{GIA} below, we show that the GIA decreases the moment of inertia, implying a negative value
of $\dot{\tau}_{\mathrm{GIA}}$. The value obtained by \citet{Mitrovica15} from new Earth's model
with a high  viscosity in the lower mantle is   $\dot{\tau}_{\mathrm{GIA}} = -1.07$ ms/cy. Thus, this
result implies,
\begin{equation}
\dot{\tau}_{\mathrm{ice+sea}} \, = \, (+1.07   - 0.45) \;\mathrm{ms/cy}  = 0.62 \; \mathrm{ms/cy} \,.
\label{ice}
\end{equation}
This result is consistent with the estimate we have in expression 
(\ref{s07}) and we apply this estimate in Table \ref{Table1}.\\

\noindent
{\it{Summary on ice melting and  sea level effects on the LOD:}}
The value of  0.7 to 0.8 mm/yr of eustatic rise of sea level for the period 1900-1990 is well supported by most authors and the IPCC.  
As a considerable deglaciation over the whole world followed the Small Ice Age of the 17$^{th}$ century, we
consider this value as a representative mean for the last three centuries. The exact value has little importance, since this
is the sum $\dot{\tau}_{\mathrm{J2}}\, =\, \dot{\tau}_{\mathrm{GIA}}+ \dot{\tau}_{\mathrm{ice+sea}}$ which intervenes
in the global estimate of the LOD and the error on the sum $\dot{J}_2$ is rather small, of about 10\%.
As derived above,  a value of  +0.62 ms/cy  may be estimated for the change of the LOD due to ice melting and  change of the sea level.

\subsection{The glacial isostatic adjustment (GIA)}  \label{GIA}

Let us now turn to the ``global isostatic adjustment'' (GIA), also previously  called    ``viscous  rebound''. 
At the end of the ice age 20 000 yr ago, the sea level was about 
125 m lower than today, the difference was kept in enormous  ice packs (up to 4 km thick) covering high latitude regions, 
see \citet{Peltier99} and \citet{Munk02}. About 7000 yr ago, the sea level was only  
2 m lower and since about 4000 yr the deglaciation is almost complete \citep{Munk02}.

The disappearance of the ice load is followed by a global isostatic adjustment. 
It is characterized by a flow, both vertically   and horizontally, of viscous matter from the Earth's interior towards  regions
previously glaciated and mainly located at high latitudes. 
The rebound reaches still today  up to 1.1  cm/yr in the SE of the Hudson Bay \citep{Peltier99}.
Such displacements are producing a decrease of the Earth's moment of inertia,
the rebound  reducing  the   Earth's oblateness. The equations of the
radial and tangential displacements as a function of latitude, longitude
and time are expressed as a function of the ice load and its spatial and time variations,  see review by \citet{Peltier04}.
The isostatic adjustment is also a function
of  the viscosity parameter varying with depth, of  the
terms in the development of the  Earth's potential and their time dependence, of the
topography, etc. Different successive models (particularly VM1 and  VM2) of viscosity  parameters
have been developed over the years. The models have a great sensitivity to the viscosity down to a depth 
of 2000 - 3000 km in the lower mantle. The influence of the Earth rotation on the GIA has been emphasized by
\citet{Mitrovica01}, who also performed models of the rebound vectors over the world.

The  models by Peltier have been  tested by 
comparing the coastal changes of the sea level over thousands of years
in various locations differently affected by the GIA    \citep{Peltier04}. The relative sea level
change (with account of the rebound) reaches up to 250 m in the South East of the Hudson Bay (with respect to 9000 yr ago), 
a similar value is observed in the Gulf of Bothnia.  These extreme changes have to be compared to the average sea level, 
which 9000 yr ago was about 12 m lower than at present, see Fig. 1 by \citet{Munk02}. The characteristic relaxation time is 
of 3400 yr in the SE of the Hudson Bay and 4200 yr in the Gulf of Bothnia \citep{Peltier04}.
The timescales of the order of several thousands of years are such that 
the contemporary deglaciation has no effect yet \citep{Sung-Ho16}, while the presently observed isostatic adjustment 
only results from the melting of the great glaciers of the Pleistocene.
By   inversion technique,  such  observations in different coastal regions provide  observational constraints 
on the viscoelastic inner structure of the Earth and on the spatial and time evolution of the glaciations. This form the 
basis of the  ICE-5G (VM2) reference model \citep{Peltier04}.

For long, there were  several rather converging estimates of the GIA contributions
to the change of the LOD. For the secular trend in the LOD, \citet{Mitrovica97}
found an acceleration of the  Earth rotation amounting to    -0.5 ms/cy on the basis of a model fitting 
the GIA decay times and the surface gravity anomalies.
The  effect of GIA on the LOD was estimated to be consistent with
-0.47 ms/cy by \citet{Johnston99} from
constraints on the sizes, locations and timing of deglaciation of the Late Pleistocene ice
sheets and current changes in polar ice caps.
\citet{Munk02}, in a critical discussion of the various terms intervening in the
LOD,  is supporting a value of -0.6 ms/cy, on the basis of a geodynamic model  \citep{Peltier98,Peltier99}, 
applied successfully to tide gauges over the world and 
which also gives consistency results for the polar wander. At that time, 
Munk considered that this value  could  account  for the difference 
between the tidal effect of 2.39 ms/cy and the historic estimate of 1.78 ms/cy given
by \citet{Stephenson16}, however the situation appears more complex.

The effects  of the GIA very much depend, as discussed above,  on the viscosity   between  the
basis of the lithosphere and the core-mantle limit \citep{Peltier04}. With a low viscosity, the timescale of the adjustment
is short and the model rapidly recovers its equilibrium. The available observations do not
seem to support such a situation.
With a  very high viscosity, the changes are damped. 
The maximum effect
corresponds to a mean viscosity. The Earth's models with standard viscosity (VM1, VM2) give a rather weak GIA effect of 
about -0.6 ms/cy, as that discussed above. The so-called MF viscosity model by \citet{Mitrovica15} is different,
with a much higher viscosity in the lower mantle.
Starting at the surface with a standard viscosity, the MF model  has a fast growing viscosity with depth, reaching a
value  of $ 2 \cdot 10^{22}$ Pa $\cdot$ s in the lower mantle, 6.3 times larger than in standard VM2 models. Such a high viscosity in the lower
mantle also corresponds to the finding  by \citet{Nakada03} and  \citet{MitrovicaF04}.
These differences in the model viscosity 
strongly change the amplitude and the time response, 
leading to a large GIA signal as well as a consistent polar wander.
These results have led \citep{Mitrovica15} to considerably revise the estimate of the  GIA effect. 
Their Fig. 3B gives a GIA contribution of $(\dot{\Omega}/\Omega) = 12.4 \cdot 10^{-11}$ yr$^{-1}$. 
The acceleration of the angular velocity 
obtained with MF model  corresponds  to a change  
of the LOD of -1.07 ms/cy, according to the expressions given in Section \ref{sealev}.
The vector expressing the polar wander obtained from the MF model is also
in agreement with observations.   
This new value  of $\dot{\tau}_{\mathrm{GIA}}$ is very different from previous estimates.
Since $\dot{\tau}_{\mathrm{J2}}\, =\, \dot{\tau}_{\mathrm{GIA}}+ \dot{\tau}_{\mathrm{ice+sea}}$ is equal to -0.45 ms/cy
from SLR observations, the above value allowed us in the last Section to ascertain the LOD effects of ice melting and sea level changes.
On the whole, the geodesic SLR data provide the most useful internal consistency check.

\noindent 
{\it{Summary on GIA effects on the LOD:}}
Most authors were supporting a moderate value of $\dot{\tau}_{\mathrm{GIA}}= -0.6$ ms/cy. 
This value of the GIA effect was difficult to reconcile  with  the geodesic results from SLR measurements.
A new  study by \citet{Mitrovica15} with a much higher viscosity in the lower mantle leads to a different value
$\dot{\tau}_{\mathrm{GIA}}= -1.07$ ms/cy. When due account is given to the ice melting and change of the
sea level, a  good agreement  with SLR geodetic observations and polar wander is obtained.

\subsection{Effects of the inner Earth's structure and core-mantle coupling}  \label{cmc}

Differential motions in or near the Earth's core may be due to 
convective motions and/or differential rotation. Such motions are 
usually thought to be the source of the Earth's magnetic field.
These motions may also operate some transfer of angular momentum
between the core and the mantle and have an effect on the LOD \citep{Hide89,Dumberry06,Souriau15}.
The inner solid core of the Earth with a radius of 1220 km is surrounded 
by a liquid outer core of radius of about 3500 km, below the solid mantle \citep{Souriau15}.

The question of a high
differential rotation between the inner and outer core has been much debated. On one side, there were 
suggestions of fast differential  rotation of the inner core
(2 or  3 degrees per year), on the other side there were more conservative studies suggesting a differential
rotation at least ten times smaller.  For example,  an  improved  analysis of waves emitted by three earthquakes 
in South Atlantic traveling through the core and analyzed with 37 seismometers in  Alaska supported  an inner core spinning only 
slightly faster (0.2 to 0.3 degrees per year) than the outer core \citep{Creager97}.
The more recent estimates from seismic observations suggest a low value of about 0.1  or less than 0.2 degree per year, 
the core turning slightly faster  than the mantle towards the East, 
see   \citet{Deuss14} and \citet{Souriau15}  for  reviews. 
A recent re-analysis of observations by the full Large Aperture Seismic Array
of  seismic waves back scattered by the inner core  for two
soviet nuclear tests of 1971 and 1974 show time shifts supporting a
differential rotation of 0.07 degree/yr, much lower than current estimates \citep{Vidale19}.
\citet{Tsuboi20} have used antipodal earthquakes and stations to analyse the
time shifts of wave  propagation  near the core. They conclude to a 
differential rotation of the inner core with respect to the crust-mantle 
of 0.05 degree/yr,  smaller than past studies, but in better agreement
with the results of \citet{Vidale19}.  

\cite{Munk60} recognized the role of core-mantle coupling 
as a source of the large  decadal LOD variations, on the basis of 
their amplitudes and timescales 
(we have seen in Sect. \ref{atm} that the exchanges with the atmosphere necessarily have a limited amplitude).  
They were followed by most geophysicists who ``supposed that
irregular ‘decade’ fluctuations in the LOD of several  ms must be manifestations of angular momentum 
exchange between the core and mantle'' \citep{Hide89,Buffett15}. 
There are, however, a few rare discordant voices, for example
\citet{Sidorenkov16}  gave arguments in favor of an origin in the  drift between the lithosphere ($\sim 60$ km) and the asthenosphere
($\sim 700$ km).
The  `decade’   fluctuations are remarkably illustrated   by the large peaks  in the variations 
of the  LOD based on lunar occultations shown by  \citet{McCarthy86}  and \citet{Stephenson16}, 
as illustrated  in  the present Figs. \ref{McC} and \ref{AV1700}. We first see a steep decrease of the LOD starting in 1860 
and reaching a narrow minimum  with - 4 ms around 1868. The full recovery occurred around 1890
and was followed by a strong  increase of the LOD with a broad maximum of about + 4 ms from 
1901 to 1912 until recovery around 1932. 
There is unfortunately no indications available on the possible cycles of such decadal oscillations.

Most remarkably, the fluid motions at the top of the core
can be deduced from the observations of the variations of the magnetic field 
at the surface of the Earth, see review by \citet{Gross09} where
many specific references can also be found. A number of authors have studied 
the  possible properties of the core magnetic field  and the following picture has emerged:
\begin{itemize}
\item  The core may be considered as a perfect conductor so that the field is frozen in it and is confined at the surface of the core by advection;
\item  The mantle is an insulator, so that  no creation or destruction of the field  occurs through it and the spatial behaviour through the mantle is predictable;  
\item The large scale flow is geostrophic, {\it{i.e.}} horizontally there is a balance between the pressure gradient and the Coriolis force; 
\item  The flow is essentially steady and has no radial component.
\end{itemize}

The first successful model able to account for the decadal LOD
variations was developed by \citet{Jault88}. They demonstrated that 
the axisymmetric components of the rotating flow in the core are
equivalently described  by the relative motions of concentric cylinders rotating around the rotation axis of the core. 
Jault et al. were able to derive the effects of the core motions on the 
LOD from the variations of the field at the surface of the core,
which were themselves obtained from the observations of the 
variations of the  magnetic field at the surface of the Earth. This finding was the opening of the connection
between the observed variations of the surface magnetic field and the changes of the LOD, a research line with  numerous papers,
see reviews by \citet{Herring07},  \citet{Olson07} and \citet{Buffett15}.
A zonal rotation of the fluid is bearing the magnetic field of the core, which is symmetric with respect to the equator 
and  mimic the  westward observed surface drift with a latitudinal dependence. A timescale 
of about 10 years characterizes this differential rotation.
\citet{Jackson93} determined the variations of the 
fluid velocities at the core-mantle boundary from data on the Earth's
magnetic field from the years 1840-1990. 
They derived the variations of the rotation of the mantle and could compare them with observations.
Several   comparisons between the observed decadal LOD variations  and the  
predictions of different models based on the observed surface 
variations of the magnetic field have been made, see for example \citet{Ponsar03}
and  \citet{Gross09}.   The agreement, although not perfect,
is good enough  to validate this remarkable connection between
what happens at the surface of the Earth and in  its deep core.

However, the way by which there is some physical transfer of angular momentum from the core to the mantle is still unclear.
The basic dynamics of Earth's core and the various mechanisms of core-mantle coupling have been reviewed and analysed
by several authors, for example by  \citet{Jault03}, \citet{Ponsar03}, \citet{Gross09}, \citet{Roberts12} and \citet{Glane18}.
Many articles and books have been devoted to the subject since fifty years, 
we  briefly try to report on the main emerging conclusions.
The different possible coupling processes  are: convection,
viscosity, magnetic field, topography and  gravity. Viscous torques appear insufficient, 
as well as convection which does not generate sufficient density  anomalies.  

The electromagnetic coupling  would be  produced by the magnetic field of the core interacting 
with electric currents induced by  the varying magnetic 
field   in the lower mantle, which
may still be weakly conducting. This mechanism can clearly make the 
necessary coupling of the mantle for producing the decadal observed change 
of the LOD \citep{Holme98,Ponsar03}. However, its efficiency  
and thus the validity of this explanation depends on the unknown conductivity, as
firstly  emphasized by \citet{Jault03},
see also \citet{Gross09}. Recently \citet{Kuang18} showed that
a different field geometry in the outer core could make the coupling 
much stronger and succeed in accounting for the polar wander.
About topography, the same kind of uncertain conclusion  is emphasized by
\citet{Jault03} and \citet{Gross09}.
There, the irregularities of the core-mantle boundary are ``pushed''
by the pressure of the liquid core and a torque is exerted on the mantle.
The size of the effect depends on the unknown amplitudes 
of the bumps on the boundary, so that topography is a valid possibility.
Another mechanism is gravity, it occurs if there are density inhomogeneities
both in the fluid core and in the mantle. Then, the local differences of gravity 
can exert a torque on the mantle. Here also, the size of the
inhomogeneities,  which determine the gravity coupling,  is unknown.
Alike the other processes, this one could transmit the core variations to the LOD, but it is also resting on uncertain parameters.

\begin{figure*}[h]
\centering 
\includegraphics[width=.75\textwidth]{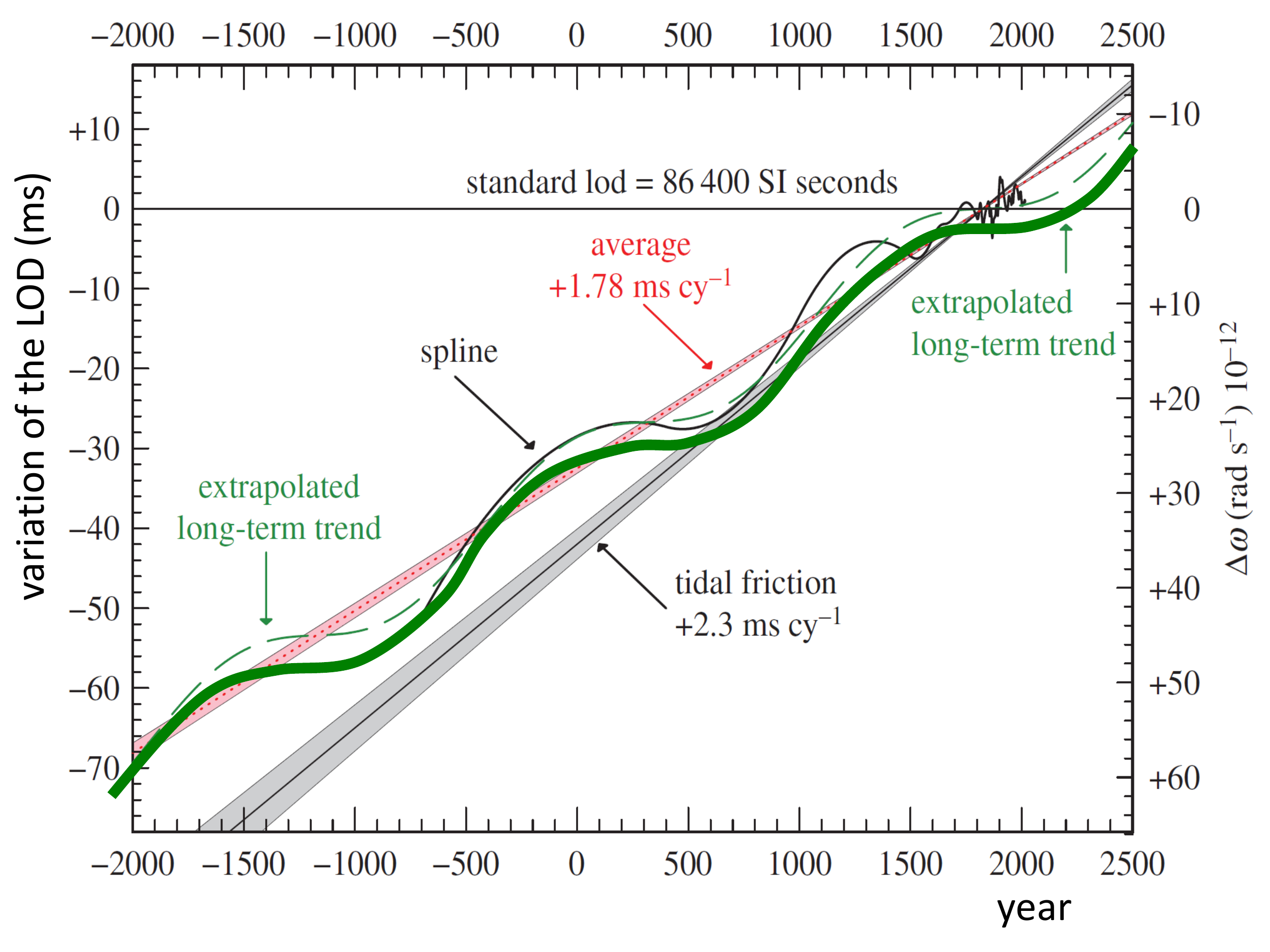}
\caption{Schematic representation of the change of the LOD over -2000 to 2500, in ms on the left axis and in angle on the right axis. 
The pink curve represents the observed average change of 1.78 ($\pm 0.03$) ms per century according to astronomical data of solar and lunar eclipses. 
The grey line shows the tidal change of 2.3 ($\pm 0.1$) ms per century from the assumed tidal friction. 
The undulated line (reinforced here) shows the extrapolated long-term trend. From Fig. 18 by \citet{Stephenson16}.
The data shows almost zero slope near year 2000 as seen in Fig. \ref{Iers}.
}
\label{ondul} 
\end{figure*}

The gravitational core-mantle coupling due to random density inhomogeneities 
in the liquid outer core and in the mantle has  been advocated
by \citet{Rubincam03} to be   the   source of the long-term 1500 yr oscillations  found by
\citet{Stephenson16} and illustrated in Fig. \ref{ondul}. A timescale of about 1000 yr corresponds
to a (rough)  estimate of the dissipation time  of the inhomogeneities.
It is most likely that this  process  resting  on gravity interactions of
random and evanescent bubbles with limited density differences  has a longer timescale than
the other ones  and could work for the 1500 yr oscillation rather than for the decadal variations.
This process could, according to \citet{Rubincam03},  also account for the westward drift
of the magnetic field by 0.2 degree/yr, while it appears to have  a negligible incidence on the polar wander. 
However, the effects on the LOD are uncertain. 

There are also many paleomagnetic data collected in lake sediments 
and old lava flows, which capture crustal features of the magnetic field. 
The records go back to about 7000 yr before present, see for example  \citet{Korte05}. 
These data lead to some modifications of the geometry and evolution of the surface magnetic field, 
which shows some  periodic eastward and westward motions of magnetic patterns. 

\citet{Dumberry06} studied with archaeomagnetic field models
the long-term behaviour of  the physical effects in the core-mantle coupling and  found significant 
differences  in the flows with respect to the case of the short-term decadal variations.
In particular, in the long-term variations there are  shears in the direction of the rotation axis, 
with a transport of angular momentum mainly parallel to this direction.
At decadal timescale, ``vertical'' oscillations dominate the transport, which  transmits the
angular momentum perpendicularly to the rotation axis.
The similarity of the period of these archaeomagnetic secular variations  to the 1500 yr  oscillations of the LOD
suggests, according to  \citet{Dumberry06}, that both variations of the magnetic field 
and the LOD originate from  angular momentum exchange between the Earth's core and the mantle. 
By inversion techniques similar to those  applied for the decadal
variations of the LOD, these authors obtained data on the  azimuthal magnetic
flows at the surface of the core over the period between 1000 BC and 1820 AD.
Thus, \citet{Dumberry06} could  calculate the exchange  of the angular momentum between the core and the mantle
and found variations at millenial-scale, consistent with the LOD variations,
thus confirming the above suggestion. As pointed out by 
\citet{Mitrovica15}, ``This prediction suggests a mean increase in Earth's rotation period over the past 2500 years as a result of core-mantle
coupling''. To estimate this increase, with account of the oscillatory
signal, \citet{Mitrovica15} performed calculations of the cumulative
$\Delta T$ for three different time series approximating  the 
change of rotation period. The average $\Delta T$ curve vs. time
they obtained has a certain thickness, representing the uncertainties due to the different approximations.
Globally, from the curve obtained by \citet{Mitrovica15} in their Fig. 3 A, we 
estimate that the secular change of the core-mantle coupling is  about the half of the shift due to the GIA, but with an opposite sign,
this gives a value of + 0.54 ms/cy, with an   uncertainty of about $\pm 0.4$ ms/cy for this
value, if we account for the various approximations performed by Mitrovica et al.
Among the various  terms in Eq.(\ref{sum}), this is the most  uncertain one, as it is based on
only one model derivation and contrary to the case of the GIA, there is no SLR measurements permitting
a check of consistency.

We note that the movements of the Earth's crust,   
the continental drifts, the plate subduction  and the Earth's cooling 
have much longer timescales, of the order of million years. 
As to the effects of   the inner structure of the Moon on the
relation between the LOD and the lunar recession, they are much smaller than those considered above.
The consequences  of a possible presence of a liquid, molten, or solid core in the Moon have been studied by 
\citet{Williams01, Shuanggen14} and the possible internal exchanges of angular momentum appear to have a negligible effect.\\

\noindent
{\it{Summary on core-mantle coupling  effects on the LOD:}}
It appears  established that the core-mantle coupling  produces variations  of the Earth's magnetic field and of
the LOD. On the basis of archaeomagnetic field models, \citet{Dumberry06} have studied the exchange of
angular momentum between the Earth's core and the mantle. Following \citet{Mitrovica15}, who support
an increase in Earth's rotation period, 
we consider a value  $ \dot{\tau}_{\mathrm{cr}}=0.54 $ ms/cy, derived from their Fig. 3 A
on the basis of  the  models by \citet{Dumberry06}. We estimate
that this value   is much more uncertain than for the other atmospheric and geophysical effects considered,
which could be checked by SLR measurements.
\\

\begin{table*}
\vspace*{0mm} 
\caption{\small 
The various contributions to the change of the LOD with and without cosmological effect on the lunar recession.
The cosmological effect of 2.75 cm/yr corresponds to the Hubble expansion with $H_0 = 70$ km s$^{-1}$ Mpc$^{-1}$, if acting
locally  in the Earth-Moon system.
Column 1 gives the  lunar recession considered of cosmological origin. 
Column 2 gives the tidal effect corresponding to the part of the lunar recession which is not of cosmological origin in ms/cy.
In column 3, only the semi-diurnal solar tide S2 is acting on long timescales.  
Columns 4-6 give  the values in ms/cy of other effects contributing  to the LOD over centuries. 
Column 7 gives $\dot{\tau}_{\mathrm{mtl}}$ which is  the total change of the LOD
as expressed by Eq. (\ref{sum}). The data in the last column have to be compared to the observed change of the  LOD,
either  to 1.09 ms/cy as determined from lunar occultations between 1680 and 2020 (Section \ref{occ}) or
to 1.78 ms/cy from eclipses in the Antiquity and Middle Ages (Section \ref{ecl}).\label{Table1} } 
\scriptsize
\tabletypesize{\squeezetable}
\begin{center} 
\begin{tabular}{ccccccc}
\hline\hline
COSMOLOGY  & TIDAL      & ATMOSPH.  & ICE MELTING & GLACIAL ISOST. & CORE-MANTLE & SUM\\
&EFFECTIVE&             & SEA LEVEL & ADJUSTMENT &         COUPLING         &$\Delta$ LOD                  \\
$(\frac{dR}{dt})_{\mathrm{cosm}}$ [{cm}/{yr}] & $\dot{\tau}_{\mathrm{tdeff}}$ [ms/cy]&$\dot{\tau}_{\mathrm{atm}}$ [ms/cy]&
$ \dot{\tau}_{\mathrm{ice-sea}}$ [ms/cy]&$\dot{\tau}_{\mathrm{GIA}}$ [ms/cy]
&$\dot{\tau}_{\mathrm{cr}}$ [ms/cy]&$\dot{\tau}_{\mathrm{mtl}}$ [ms/cy] \\
\multicolumn {7} {l}{ $\quad \quad \quad  \quad  \quad  \quad \quad \quad \quad \quad \quad \quad \quad \quad \quad \quad \quad \quad
\quad \quad \quad \quad \quad \quad \quad \; \mathrm{Sum\; of\;ice\;melting+sea\;level+GIA}$ }   \\  
\multicolumn {7} {l}{ $\quad \quad \quad  \quad  \quad  \quad \quad \quad \quad \quad \quad \quad \quad \quad \quad \quad \quad \quad
\quad \quad \quad \quad \quad \quad \quad \quad \; \mathrm{from\; Satellite\;Laser\;Ranging}$ }   \\  
\hline\hline
\underline{No cosmol. effect:} & &  & & & & \\
0  cm/yr  &  2.395   &      -0.05 $(\pm0.02)$    &    0.62    & -1.07   &      0.54 $(\pm0.4)$   &   2.44 $(\pm0.4)$        \\
\multicolumn {7} {l}{ $\quad \quad \quad  \quad  \quad  \quad \quad \quad \quad \quad \quad \quad \quad \quad \quad \quad \quad \quad
\quad \quad \quad \quad \quad \quad \quad \quad \quad \quad \quad  \; -0.45 (\pm0.05)\; $ }   \\  
\hline
\underline{With cosmol. effect:} & &  & & & & \\
2.75 cm/yr    &  0.68   &      -0.05 $(\pm0.02)$   &    0.62   & -1.07  &      0.54 $(\pm0.4)$  &    0.72  $(\pm0.4)$     \\     
\multicolumn {7} {l}{ $\quad \quad \quad  \quad  \quad  \quad \quad \quad \quad \quad \quad \quad \quad \quad \quad \quad \quad \quad
\quad \quad \quad \quad \quad \quad \quad \quad \quad \quad \quad  \; -0.45\; (\pm0.05)$ }   \\  
\hline\hline
\end{tabular}
\end{center}
\normalsize
\end{table*}

Table \ref{Table1} summarizes the change rates of the  LOD for the various effects examined.
The case without  a cosmological contribution is presented in block 1, with lines 1 and 2. 
In block 2, with lines 3 and 4, 
it is assumed that the Hubble expansion with $H_0=70$ km s$^{-1}$ Mpc$^{-1}$ is also present
in the Earth-Moon system. This corresponds to a lunar recession of 2.75 cm/yr. The difference with observed
lunar recession of 3.83 cm/yr is   $\left({dR}/{dt}\right)_{\mathrm{tdeff}}= 1.08$ cm/yr, 
thus   assumed to originate only  from the tidal effect, see Eq. (\ref{TR}).  
The corresponding tidal slowing down of the Earth rotation  would   thus be 0.68 ms/cy instead of 2.395 ms/cy.   
Interestingly enough, the sum of the various positive and negative contributions (\ref{sum})
other than the tidal one appears to be rather small of about +0.04 ms/cy.
When all   contributions to the LOD
are accounted for, one obtains the results in the last column of Table \ref{Table1}. In the case without a
cosmological effect, a slowing down of 2.34 ms/cy is obtained, while it is 0.62 ms/cy  if the Hubble expansion
is present. These values are to be compared to observations.

\section{Survey of data  sources on the variations of the LOD}  \label{LOD}

\begin{table*}[h]  
\normalsize
\vspace*{0mm} 
\caption{\small
The different estimates of  the secular increase of the LOD. \label{Table2}}
\begin{center} 
\begin{tabular}{ccc}
\hline\hline
Source and period of observations &  Change of the   LOD  &References\\
&      ms/cy &      \\
\hline
&   &      \\        
Ancient eclipses    700 BC - 1600 AD    &  $1.40 $ &   \citet{Stephenson84}   \\
Comparison of ephemeris 1875-1970    & $2.49 $&   \citet{Goldstein85}    \\
Lunar occultations  1656-1986            &  $0.73  $  &  \citet{McCarthy86}   \\ 
Ancient eclipses   700 BC - 1600 AD    &  $1.70 $ &   \citet{Stephenson95}  \\ 
Combined data    1832.5 -1997.5          &                 --                   &    \citet{Gross02}    \\
Lunar occultations over last 350 yr     &  $0.90 $ &  \citet{Sidorenkov05}          \\
Ancient eclipses   720 BC - 1600 AD    &  $1.78 $ &  \citet{Stephenson16}          \\
Paleontology   $\sim$ 1 Gyr            &      $ 1.64   $ & \citet{Deines16} \\
Sea level change since 3000 yr      &          $ 0.00    $ &   \citet{Hay16} \\ 
Babylonian obs. 419 BC -89 BC        &                    --                     & \citet{Gonzalez18} \\
Lunar occultations+IERS  1680 -2020   &        $ 1.09   $                &    This work \\
\hline
\end{tabular}
\end{center}
\end{table*}

There are different sources of information on the changes of the LOD: 
\begin{itemize}
\item The study of the ancient solar and lunar eclipses over the last 2700 yr; 
\item The analysis of lunar occultations of  bright stars and planets,  with precise times, since the middle of the 17$^{\mathrm{th}}$ century;
\item The variations of the sea level;
\item The precise geodetic measurements  of the Earth rotation since 1962 with the IERS.
\end{itemize}

\subsection{Ancient observations of eclipses}  \label{ecl}

The study of ancient lunar and solar eclipses offers
a powerful test,  since  large  cumulative shifts $\Delta T$ may result between the time UT dependent on Earth rotation and TT 
based on constant days of 86400 SI seconds.  For example, over an interval of 2000 yr, the shift of the
timescale is  of the order of 18 000 s. Such differences correspond 
to displacements of the predicted  geographic locations of the eclipse
visibility over large distances, up to a few thousand km.
\citet{Stephenson84} determined from eclipse data from 700 BC to 1980 AD
a change of the LOD by $1.4 \cdot 10^{-5}$ s $\cdot$  yr$^{-1}$.
Later  \citep{Stephenson95} corrected their previous value and gave   $1.70 \cdot 10^{-5}$ s $\cdot$  yr$^{-1}$.
A further  study by  \citet{Stephenson16}  lead to 1.78 ms/cy  from the data 
prior to AD 1600, a work  often  used as a comparison basis for the LLR results. This value
is much lower than the value of 2.395 ms/cy   \citep{Williams16a,Williams16b} 
for   a tidal interaction producing the observed 3.83 cm/yr lunar recession, see also \citet{Stephenson20}
who consider the difference as significant.

In the works of 1995 and 2016, Stephenson et al.  found an 
undulation with an amplitude of about 4 ms around the mean with a period of about 1500 yr, so that at some epochs
the centennial increase of the LOD is higher than the 
mean and lower at other epochs, see Fig. \ref{ondul}  derived  from  Fig. 18 by \citet{Stephenson16}.   
For example, the average increase would be     lower between about AD 1600 and 2100. 
This long-term  undulation has often been a source of stimulation and  a comparison basis for the
studies of the core-mantle coupling, as seen in the previous Section. However, as noted by \citet{Stephenson16} about
the 1500 yr undulation, ``this is less reliably established because it is critically dependent on relatively few observations.''

\citet{Dalmau97} criticized    the results  by \citet{Stephenson95}, 
in particular the treatment of the Arabic records and the statistical method employed:
``the authors listed above have not evaluated the Arabic records with the sufficient care. 
They have taken the numerical data from the sources without considering the context. The
statistical method used by them meets by no means the requirements of the subject.''
\citet{Hay16} have pointed out that the inferred departure from a quadratic relation proposed by Stephenson et al. 
is driven by a relatively small number  (11) of solar eclipses untimed observations (without indications of the eclipse duration).
\citet{Soma16} have used two ancient Japanese occultations of Venus
in AD 503 and of Saturn in AD 513
to test the secular change of the LOD. For both observations, they
found a $\Delta  T$ much smaller,  with a difference between about 400 s and 2700 s,
than the smoothed value of \citet{Stephenson97}. 
This difference is supported by the data from solar eclipses in AD 516 and 522. 
Their conclusion was that in the beginning of the sixth century AD, the $\Delta T$ values were well below the smoothed
trend obtained by \citet{Stephenson97}. 
Another eclipse of AD 454 showed a higher $\Delta T$ at that time, suggesting fast variations.
The Babylonian occultations and appulses between 80 and 419  BC were studied  by \citet{Gonzalez18}, 
who suggested a small revised version of the quadratic fit  with respect to \citet{Stephenson16}.
Thus, despite their high interest the ancient observations contain  sources of  uncertainties, 
which may affect the conclusions. A new independent study would be most useful in the context.

\subsection{Previous studies from  lunar occultations}

From AD 1623 up to 2015 there were
478'843 observations of timed  occultations of stars by the Moon collected by 
\citet{Herald12} in CDS at Strasbourg, who provided a most reliable sample.
Before the middle of the 17$^{\mathrm{th}}$ century, the timing of the occultations was  often inaccurate 
and the observations show a large scatter.
\citet{McCarthy86} have made a detailed study of the evolution of the LOD
for the period of AD 1656 to 1986. The observations come from four different sources, 
one of them is McCarthy himself observing
at the US Naval Observatory photographic zenith tube. All   
data reduced to the same timescale, with the  values of the LOD and their errors, are
given in the paper. Some light smoothing has been applied, particularly for the data before 1820. 
Over the whole period, \citet{McCarthy86} found
a mean decrease of the angular velocity of the Earth of 
\begin{equation}
\frac{\dot{\Omega}}{\Omega} =
-8.433 (\pm0.207) \cdot 10^{-11} yr^{-1} \, .
\label{sec}
\end{equation}
Converted in term of the LOD, this braking corresponds to  a
change of +0.73 $(\pm 0.018)$ ms/cy. The curve obtained by \citet{McCarthy86} is illustrated in Fig. \ref{McC},
where the authors have shown only the deviations around  the mean value given by the above equation.
We see that over an interval of 330 yr  the mean of the data are very well represented by a horizontal line.
One could wonder whether  the decline appearing near the right end of the plot (year 1986)
is further going on and supporting the general  flat curve.
This is  the case as illustrated by Fig. \ref{Iers} of the IERS data, 
which shows that the decline after  1986 is going on with  further oscillations
around  relatively low LOD values.

The \citet{McCarthy86}  data over the last 3.3 centuries were supporting  a secular change of the LOD 
about 2.4 times smaller than the values from antique observations.
Commenting on this difference,  \citet{McCarthy86} noted the possibility 
that errors in the ancient observations have led to systematic errors by Stephenson and Morrison. Another possibility is that 
the secular lengthening since 1656 is different from the overall trend from 700 BC to 1600 AD. However, it is  surprising that
this  large difference in the  LOD evolution  should coincide with the appearance of accurate time determinations
and telescopic observations.

From astronomical telescopic observations over the last 350 yr, 
\citet{Sidorenkov05} found a  value of $0.9\cdot10^{-5}$ s $\cdot$yr$^{-1}$ for the secular change of the LOD.
This result is   not very different of that by \citet{McCarthy86}.
The  value  by Sidorenkov  has been first 
used by  \citet{Dumin05} to study the possibility of cosmological expansion in the Solar System  due 
to Dark-Energy effects and then in further studies on local expansion  \citep{Dumin07, Dumin16}. 
A combined LOD series spanning the years 1832-1997 has been
established by \citet{Gross02}. It is based on lunar occultations, optical astronomy, satellite laser ranging and VLBI.  Over the period
considered, the observations are  in excellent agreement with \citet{McCarthy86}
and \citet{Stephenson84}. Table \ref{Table2} collects the different estimates of the secular variation of the  LOD.

The evolution of the LOD,   derived from high quality lunar occultations since about 1700,
is shown  in Fig. 19 by \citet{Stephenson16}. 
This figure is given in Fig. \ref{AV1700} below.
It confirms the relatively low change of the LOD over the last centuries in agreement
with the results by \citet{McCarthy86} and \citet{Sidorenkov05}.
Many oscillations of various characteristic times are
visible in this figure, in particular the so-called decadal fluctuations with large peak-to-peak amplitudes
with a strong one between the years 1860 and 1935, see Sect. \ref{cmc}.

\subsection{New study of the LOD from  lunar occultations}  \label{occ}

Some time having passed since the work by \citet{McCarthy86}, it is useful to make another 
study of the LOD covering the recent centuries up to the present.  
We take the tabulated data of lunar occultations  by \citet{Stephenson16} 
from 1680 and extend them to 2020 thanks to the recent  IERS values (see the IERS URL to the DataProducts/EarthOrientationData/eop.html).
The thin black curve in  Fig. \ref{AV1700} illustrates the sequence  of
lunar occultations given by  Stephenson et al.  The other thin curves show  other  data (see caption). 
In their common parts, they agree well.

A linear regression of the LOD based on data from the year  1680 to 2020  gives  the following expression,
\begin{equation}
\mathrm{LOD [ms]} = 0.01089 \, \cdot  \, \mathrm{year [AD]} -20.068
\label{Eqlod}
\end{equation}
As seen from (\ref{Eqlod}), the zero of LOD is in years $1842-1843$ in agreement with Fig. \ref{ondul}.
The zero in this case corresponds to a day of about 86400 SI seconds exactly. 
While, the secular change of the LOD  given by the above relation is:
\begin{equation} 
\mathrm{Slope \;of\; the\; LOD}  = \, 1.09 \; (\pm 0.012) \; \mathrm{ms/cy} \,,
\label{Delta}
\end{equation}
where the 90\% confidence limit is indicated.
The regression line given by Eq. (\ref{Eqlod}) is illustrated by the thick black line in Fig. \ref{AV1700}. 
The main deviations from the mean trend  in  the observations shown in this  figure 
are  similar to those of Fig. \ref{McC}  by \citet{McCarthy86} except that 
unlike in  Fig. \ref{McC} no adjustment for the mean value have been applied.
The present result on the change of the LOD is supporting a relatively low value, relatively close to the result 
 by \citet{McCarthy86} and by \citet{Sidorenkov05}, despite
the differences in  the start  and the end  of the periods considered. 

In the calculations for (\ref{Delta}), the tabulated data of lunar occultations by \citet{Stephenson16} was utilized.
To obtain the change in the LOD \citet{Stephenson16} applied cubic splines fit to their data about the time change $\Delta$T.
Where $\Delta \rm{T}= TT-UT$ is defined as the difference between the theoretically uniform time scale, 
which is denoted by Terrestrial Time (TT), and the (variable) rotational period of the Earth denoted by Universal Time (UT).
The cubic spline fit allowed for evaluating the time derivative of $\Delta$T and thus estimating the change in the LOD, 
which was provided in their tabulated results. To validate the values obtained from the cubic spline fit method they 
utilized also local polynomial regression (loess) with a quadratic smoothing parameter throughout after 1600.
The results based on the cubic spline fit and loess can be seen in the original Fig. 19 of \citet{Stephenson16},
as well as in Fig. \ref{AV1700} here. \citet{Stephenson16}  have also used parabolic fit to a larger data set 
$-720$ to 2015 with the unit of the second on the Ephemeris Time (ET) scale determined from the observations 
between 1750 and 1892 and were carefully discussed in the computation of $\Delta$T \citep{Stephenson16}  
along with the Lunar ephemeris and various relevant observations and comparisons.

\begin{figure*}[h]
\centering
\includegraphics[width=.75\textwidth]{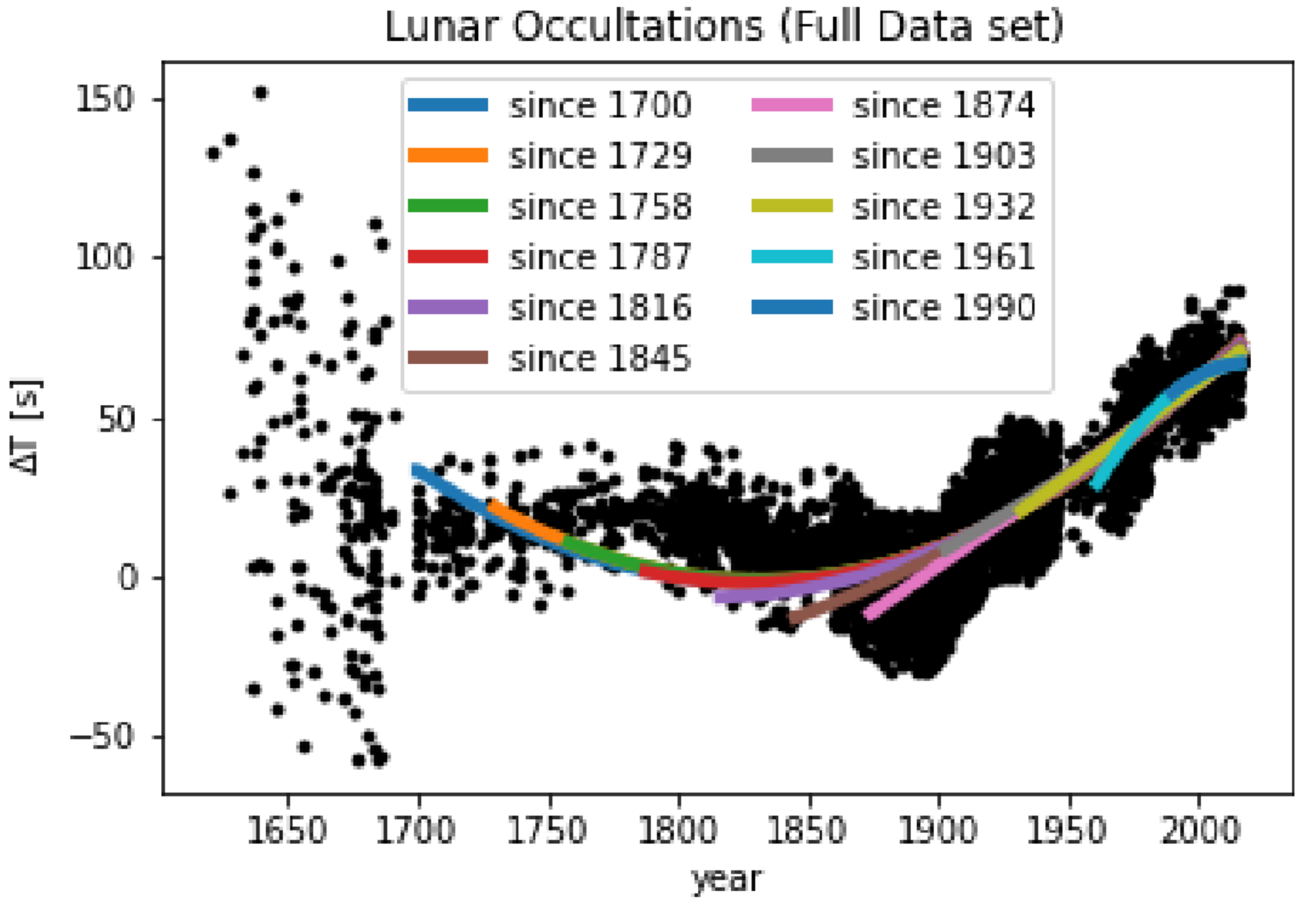}
\caption{Lunar Occultation data on $\Delta$T 
from the original data provided in the supplemental material of \citet{Stephenson16}.
The corresponding quadratic polynomial fit using the records as indicated are shown.}
\label{Lunar Occultation} 
\end{figure*}

\begin{figure*}[h]
\centering 
\includegraphics[width=.75\textwidth]{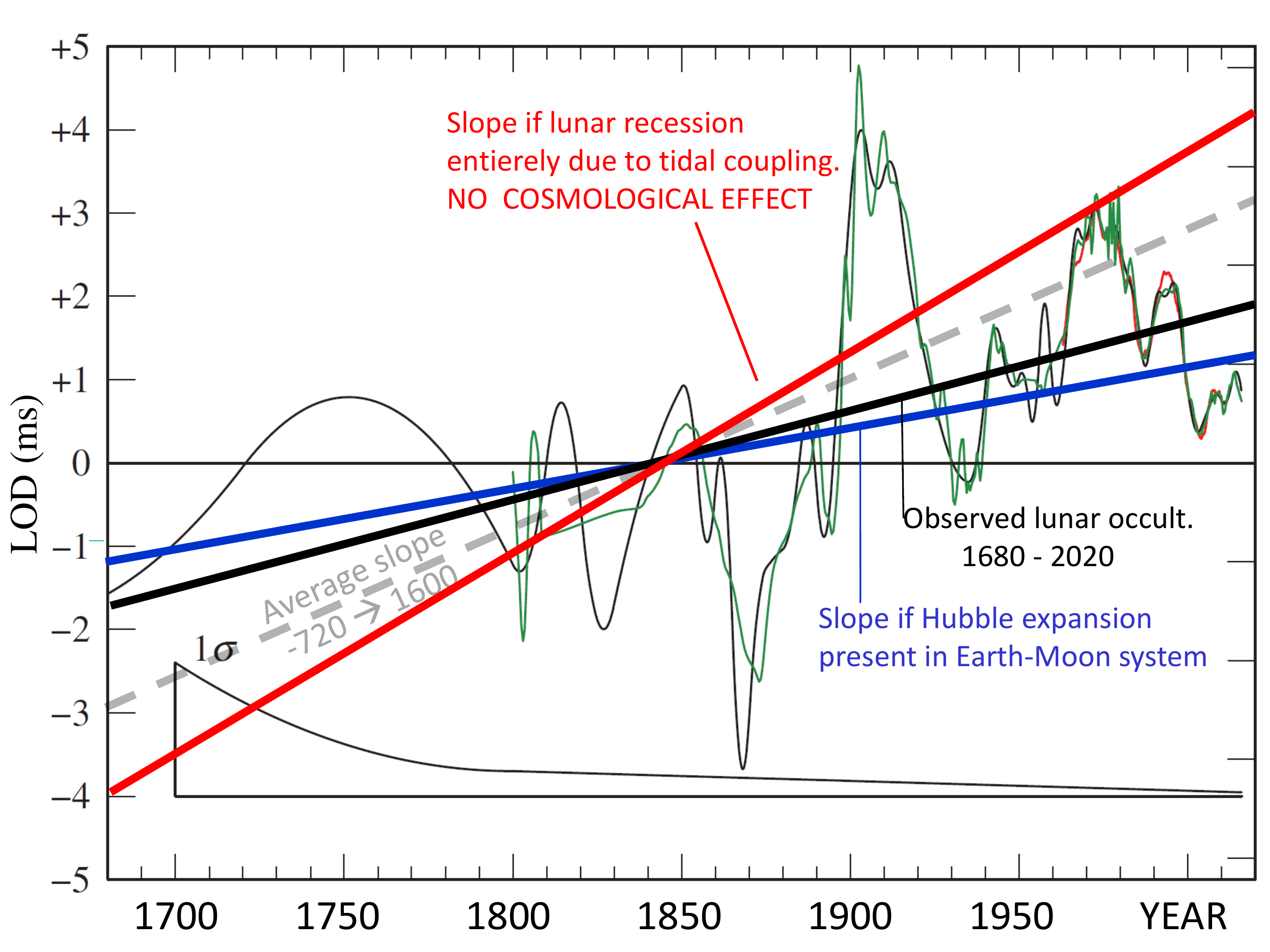}
\caption{
The change of LOD from the lunar occultations between 1700 and 2015  according to Fig. 19 by
\citet{Stephenson16}. The thin black observed curve is a spline fitting, the thin green
curve after 1800 is obtained from a local moving  regression (loess)
and the thin red curve shows the IERS data (cf. Fig. \ref{Iers}). 
The thin lines belong to the original figure  by Stephenson et al., while the thick lines are added in the present work.
The thick black line is the regression line of the observations of lunar occultations  obtained from the tabulated data by 
\citet{Stephenson16} completed by recent IERS data. It crosses the value zero for the  change of LOD in the year 1842.
The other  thick lines show different slopes, also  attached to a  LOD change of 0 ms in the year 1842.
The thick red line indicates the slope of a change of the LOD
by  2.395 ms/cy  \citep{Williams16a}, 
obtained if the lunar recession is entirely due to tidal coupling of the Earth-Moon system. 
Closely above this value would, we would have the slope of 2.44 ms/cy corresponding to  
the case of no cosmological effect in the lunar recession (cf. Table \ref{Table1}).
The thick grey line  gives  the 
slope of 1.78 cm/cy  obtained from ancient eclipses from 720 BC to 1600 AD by \citet{Stephenson16}.
The blue line shows the slope of 0.72 ms/cy of the LOD  if the Hubble expansion participates  (2.75 cm/yr) 
to the observed lunar recession (cf. Table \ref{Table1}).}
\label{AV1700} 
\end{figure*}

In order to identify the size of the data needed to detect the long term impact in the change of the LOD,
we have used the original records by \citet{Stephenson16} since year 2020 with backwards time increments of 29 years 
and have utilized a polynomial of second degree  for our least square fit. 
The corresponding best fit curves are shown in Fig. \ref{Lunar Occultation}.
The resulting coefficients are shown in Table \ref{polynomial} along with the relevant slope value.
As seen from the table, after 200 years of data the change of the LOD settles at the
value consistent with the one computed in (\ref{Delta}) based on the cubic spline fit tabulated results.
Thus, in order for the effect to be confirmed and to become visible within the highly accurate modern IERS data, 
another 150 years of collection of new data may be needed. 


\begin{table}[h]
\normalsize
\vspace*{0mm} 
\caption{\small
Values of the LOD change (slope) using data sets starting since the indicated year and ending in 2020.
The slope value is given by the expression $2\times100\times365\times(10^{-6}\times\,a_2$). 
The $a_i$ are the coefficients deduced via quadratic least-squire fit to $\Delta$T as function of the Julian day.
The $\chi$ is the square root of the residuals per data sample adjusted for the degrees of freedom.\label{polynomial} }
\begin{center}
\scriptsize
\begin{tabular}{||c|c|c|c|c|c||}
\hline\hline
record &  slope &  $a_2$ &  $a_1$ &         $a_0$ &    $\chi$  \\
since &  [ms/cy] &  [ns/day$^2$] &  [s/day] &    [s] &     \\
\hline
 1700 &        1.1681 &      16.0020 &   -0.0765 &   91316 & 4.8264 \\
 1729 &        1.1988 &      16.4222 &   -0.0785 &   93804 & 4.8108 \\
 1758 &        1.1983 &      16.4147 &   -0.0785 &   93759 & 4.8058 \\
 1787 &        1.1651 &      15.9596 &   -0.0762 &   91061 & 4.7866 \\
 1816 &        1.0195 &      13.9659 &   -0.0665 &   79229 & 4.6512 \\
 1845 &        0.6348 &       8.6963 &   -0.0408 &   47925 & 4.2468 \\
 1874 &        0.0177 &       0.2426 &    0.0004 &   -2362 & 3.4032 \\
 1903 &        0.4243 &       5.8124 &   -0.0268 &   30890 & 2.8724 \\
 1932 &        0.0639 &       0.8756 &   -0.0027 &    1299 & 2.3726 \\
 1961 &       -6.0413 &     -82.7581 &    0.4070 & -500217 & 1.3003 \\
 1990 &       -6.1537 &     -84.2970 &    0.4145 & -509459 & 1.0348 \\
\hline
\end{tabular}
\end{center}
\end{table}

The difference between  the above  results and the data from ancient eclipses between 720 BC and 1600 AD is quite large.  
Both sampling have their respective advantages and weaknesses. The eclipse data are much less accurate,
but they  cover a much longer period of time. The occultation data are much more accurate, but they cover a shorter period.
As already pointed out by \citet{McCarthy86}, there are two possibilities.
A)  ``It is possible that errors in the ancient observations have resulted in a systematic error in the estimate of Stephenson and Morrison''. 
This possibility could be supported by some remarks  by  \citet{Dalmau97} and \citet{Hay16}.
Also,  it is bit surprising that this change of LOD regime just corresponds to the epoch where timed observations started.
B) The difference is real and represents a systematic  change of the LOD.  The 1500 yr oscillation, if real, might be an explanation. 
(This 1500 yr ondulation has been considered very seriously by geophysicists studying the core-mantle
coupling, see Sect. \ref{cmc}).

In Fig. \ref{ondul},  we see that  from 720 BC to to 1600 AD, there are two phases where the increase rate of the LOD
is high (from  about 700 BC to 100 BC, and from about 900 AD to 1600 AD), 
while there is only one  phase (from about 100 AD to 700 AD) where the rate is shorter than the mean of 1.73 ms/cy.  
This difference is  biasing the mean increase rate of the LOD by \citet{Stephenson16}  towards a high value, 
since the mean  rate is the mean slope of the cumulative effect   $\Delta T$ vs. time.
We may call this the ``high-average regime''. The period from about 1600 AD to 2100 AD lies in the relatively ``flat  regime'' of the undulation, 
where the increase rate of the LOD is lower than the mean. Thus, this could explain the  difference of  the results by \citet{Stephenson16} 
with those of \citet{McCarthy86} and ours. In this respect, we note that  the period of 1970 to 2015, 
where the LLR observations have been made, is also fully lying in such flat regime, which may account for its peculiarity.

\subsection{The LOD from the IERS.  Other tests} 

The most accurate  information on the contemporary variations of the LOD   
is provided by  the International Earth Rotation Service  (IERS).
The studies are  based on  ground based  observations, geodetic satellites 
and VLBI observations of most distant objects to ascertain the various geodesic 
parameters and the LOD in particular \citep{Herring00}. 
The consistency of recent data for Earth rotation and gravity field parameters obtained from 
Satellite Laser Ranging (SLR) has also been demonstrated \citep{Sosnik16}. 
Fig. \ref{Iers} illustrates the LOD, the excess revolution time, from 1962 to 2020.
The decadal fluctuations in this interval prevent, for the time being, 
any derivation of a mean trend on such a small period of time. 

It is worth noticing that the IERS data shows slight speeding of the earth rotation 
resulting in a shorter day of about $0.0001$ ms/yr,
which is easily noticeable by eye inspection of the data in Fig. \ref{Iers}.
Since the melting of glaciers at moderate latitudes and polar ice-fields 
has no effect anymore and actually results in the opposite sign 
of the LOD change, as discussed in \ref{sealev},
it is an interesting question for mathematical modeling if this numerical value of $1\cdot10^{-7}\rm{s}\cdot\rm{yr}^{-1}$
could be explained by the disappearing of the mountaintop glacier ice in the tropics 
around the world due to the observed accelerated environmental changes \citep{Thompson21}, or if this is
due to global warming effects in the semi-diurnal atmospheric tide S2 which has the same-sign
effect on the LOD change as discussed in section \ref{atm}.

We note that  \citet{Goldstein85} gave a high value $2.49\cdot10^{-5}\rm{s}\cdot\rm{yr}^{-1}$ 
for the change of the LOD over the period 1875-1970 on the basis 
of comparisons between the longitudes and the  mean motions of the Sun from ephemeris tables. However, the time period is too
short  to give a significant trend, although the order of magnitude is correct. \citet{Deines16} determined the average deceleration of the  
rotation rate of the Earth from paleontological fossils and deposits
back to more than a Gyr ago. They obtain a deceleration rate equal to 96.6\% of that of \citet{Stephenson95}, this
corresponds to 92.3\% of the rate by \citet{Stephenson16}, which is a bit surprising since the Moon was closer to the Earth
and the tidal effects larger. 

\citet{Hay16} use the sea level changes over the last 3000 yr to infer the change of the LOD, accounting for the ice mass fluctuations, 
the thermal variations, the change of shore lines, the viscoelastic response of the mantle,  etc.
They find large differences in the change of the LOD with respect to the eclipse predictions, 
with a curve in sawtooth and no  clear global trend. 
As a matter  of fact, the curve they obtain for the secular variations of the
LOD is compatible with  0 ms/cy or even with a marginal acceleration of the Earth's rotation. However,
we may remark that the inferences on the LOD from the evolution of the sea levels is a very  convoluted test.

\section {Analysis of the LLR and LOD relations}  \label{disc}

We now compare these various predictions and observations. The first question is which observations to
consider, the ancient eclipses or the lunar occultations.  

\subsection{The ``high-average'' and the ``flat'' regime}  \label{regi}

The high mean value of the Earth's deceleration from eclipses  between 720 BC and 1600 AD,  
and the low value from lunar occultations between 1680 and  2020, if not due to errors, 
could correspond to  different regimes within the 1500 yr oscillation: the ``high-average regime''
and the ``flat regime'' respectively, as discussed previously (Sect. \ref{occ}).
This oscillation  may result from angular momentum exchange 
between the core and the mantle, as proposed by some geophysicists and  discussed in  Sect. \ref{cmc}.
But, the reality of the oscillation is uncertain and the same for the origin of the difference
between the two regimes.

An exchange of angular momentum between the core and the mantle  lets the total Earth's angular momentum
unmodified, so that there is  no direct effect of the exchange on the lunar recession.
The  change $\Delta LOD$ in the mantle only modifies insignificantly $T^2_i$ in the term $k_i$ so that the
effect of the  tidal coupling remains  essentially the same.

However,  the velocity of lunar recession of 3.83 cm/yr is not an inertial velocity: if  the gravitational pull exerted by the
tidal deformation would suddenly be absent (or different),  about  one second  later there would no longer 
be any increment (or a different one) in the Earth-Moon distance.  The tidal coupling is a gravitational effect, as such it
is  acting  at the speed of light. As the origin of the different regimes is highly uncertain (we have noted that
the difference appears with the epoch of accurate observations), LLR and LOD observations in the same epoch
may be an advantage.

\subsection{Comparisons with different solutions}

Fig. \ref{AV1700} presents graphically the main results of this work. The thick
black straight line represents  the mean of the  observations of lunar occultations + IERS data, as described 
by Eq. (\ref{Delta}) with a slope of 1.09 ms/cy. 
Three  other slopes are shown in Fig. \ref{AV1700}, also attached in the year 1842 to a LOD change equal to zero.
The first one in red color represents a change of  2.395 ms/cy,   corresponding to the case where    the whole observed 
lunar recession would be due to the tidal effect \citep{Williams16a}. This red  line  is  steeper by a factor 2.2
than the general observed trend over  340 yr, this is quite a large difference, larger than expected from
a combination of the  tidal and the geophysical effects 
studied in Sect. \ref{phys}. Thus,  the low slope observed over 340 yr   makes it difficult 
to support that the whole lunar recession is due to tidal effect. 
We  also see from Table \ref{Table1} that if no cosmological
effect is present, accounting for all corrections,  we would have a slope of 2.44 ms/cy      
(which would be slightly steeper than the red line).
This last solution and the  red  line are also  not in agreement with
the grey  line giving the slope from eclipses in the Antiquity and Middle Age, 
despite all the geophysical corrections applied to the LOD.
Thus, there may still be some unknown effect  accelerating the Earth rotation, 
or there may be another significant expansion  present in the observed lunar recession, whatever data is considered.

The grey broken line in Fig. \ref{AV1700} represents the slope of 1.78 ms/cy obtained  with the LOD data from 720 BC to 1600 AD by
\citet{Stephenson16}.  It is certainly a  ``high-average'' value
since the period considered  contains two phases of  fast change, if the oscillations of the LOD are real.
As noted above, it is in disagreement with both the results with and without a cosmological effect,  
the difference with respect to the results of the last 340 yr being larger.

We now examine   the blue slope  in Fig. \ref{AV1700}. It corresponds to  the third line in Table \ref{Table1}, 
where it is assumed that the Hubble-Lema\^{i}tre expansion
with $H_0 = 70$ km s$^{-1}$ Mpc$^{-1}$ is acting in the Solar System. This means an expansion of  2.75 cm/yr
of cosmological origin in the Earth-Moon system  among  the observed  3.83 cm/yr. In this case, a recession of 
only  1.08 cm/yr would be  due to the tidal interaction with the Earth. Wether such a low tidal interaction would be
compatible with the Earth's oblatness  remains  an open question.
The conservation of the angular momentum would thus imply a change of the LOD of 0.68 ms/cy. 
With account  of the various recent estimates of the geophysical effects (cf. Table \ref{Table1}),  
we find a  resulting change of the LOD of 0.72 ms/cy.
This value is  close to that given by the observations of lunar occultations (1.09  ms/cy) 
from 1680 to 2020 (black line in Fig. \ref{AV1700}).

In the above considerations, we have used the Hubble constants at the local and cosmic scales to be the same. 
However, in principle, it could be conceived that the local Hubble constant might be somewhat
different from the one at the global scale (e.g., by $10-20\%$), 
for more details, the reader is referred to the works by \cite{Dumin08, Krizek15, Dumin18, Krizek21}.
Allowing such freedom in the value of the local Hubble expansion could be utilized to fit the data better.

On the whole, it  appears that  the LOD and LLR data  may admit
the occurrence of an additional contribution to the lunar recession, of the order the cosmological expansion,
if we consider the results from the lunar occultations over the last 340 years, alternatively there may 
be an unknown source of  acceleration of the Earth's rotation.
Also, this is consistent with the fact that  the physical action of the tidal forces on the Earth's mantle, hydrosphere, 
and atmosphere can only account for about 55\% of the observed lunar recession \citep{Krizek15}.

\section{Conclusions}  \label{concl}

The observations of  lunar occultations completed by the IERS data show for  the period from 1680  to 2020 AD
a variation rate of the LOD equal to 1.09 ms/cy. This rate is in agreement with the results of  \citet{McCarthy86}
and \citet{Sidorenkov05}. The above rate is
 lower than the mean of 1.78 ms/cy derived on the basis of the data for eclipses from the  Antiquity  to 1600 AD 
by \citet{Stephenson16}. We also noticed that the difference
in the two rates appear at the epoch of a major  change in the data accuracy with telescopic observations.

The comparison made in Fig. \ref{AV1700} 
shows first  that the hypothesis of a lunar recession of 3.83 cm/yr entirely due to  tidal effects 
meets some difficulty with the LOD observations, whether from eclipses in the Antiquity (as also mentioned by \citet{Stephenson20})
or from lunar occultations over the last 340 years.
Secondly, when due account is also  given to the   recent detailed 
studies on the atmospheric effects,  the  melting from icefields, the changes of the sea level, the 
glacial isostatic adjustment and the core-mantle coupling,  the agreement between the LOD predictions 
and observations  is equal or better  for the case of lunar occultations  and the occurrence of a cosmological expansion, than
for the case of antique eclipses and no cosmological effect (cf. Table \ref{Table1}). This is in agreement with some results by 
\citet{Dumin07,Dumin08,Dumin16}.

The present study clearly shows how rich and convoluted is the subject of the LOD variations and that
there are still uncertainties in geophysical processes, in particular concerning the core-mantle coupling.
Thus, we   do not claim this work gives  a proof of the cosmological expansion at small scales,  but we 
may say  that this old question is still open.


\acknowledgments
Thanks are  expressed to Dr. Wolfgang Dick  for providing information on the LOD from the IERS.
A. M. expresses his deep gratitude to his wife and to D. Gachet for their continuous support.
V. G. is extremely grateful to his wife and daughters for their understanding and family support. 
We are particularly grateful and highly appreciate the in-depth reading and careful reviewing of our paper by the referee,
who has made many remarks that have improved the quality of the paper. 

\section*{Additional statements}

\begin{fundinginformation}
This research received no external funding.
\end{fundinginformation}
\begin{dataavailability}
The datasets underlaying this article were derived from sources in the public domain:\\
\small
{\href{https://www.iers.org/IERS/EN/DataProducts/EarthOrientationData/eop.html}
{\detokenize{www.iers.org/IERS/EN/DataProducts/EarthOrientationData/eop.html}}}\\
\url{https://www.iers.org/IERS/EN/Science/EarthRotation}\\
\url{https://astro.ukho.gov.uk/nao/lvm}\\
\normalsize
\end{dataavailability}

\begin{ethics}
The authors declare no conflict of interest.\\
\end{ethics}

\end{document}